\RequirePackage{fix-cm}
\documentclass{svjour3}                     % onecolumn (standard format)
\smartqed  % flush right qed marks, e.g. at end of proof
\pdfoutput=1
\usepackage{graphicx}
\usepackage{longtable,tabularx}
\usepackage[english]{babel}
\usepackage{subfigure}
\usepackage{amssymb}
\usepackage{amsmath}
\usepackage{multirow}
\usepackage{booktabs}
\usepackage{graphicx}
\usepackage{epstopdf}
\usepackage{pdflscape}
\usepackage[english]{babel}
\usepackage{subfigure}
\usepackage{threeparttable}
\usepackage{enumerate,graphicx,relsize}
\usepackage{longtable}
\usepackage[toc,title,page]{appendix}

\usepackage{epstopdf}

\usepackage{tabularx}
\usepackage{threeparttable}
\usepackage[figuresright]{rotating}
\usepackage{xcolor}
\usepackage[margin=2cm]{geometry}

\usepackage[misc,geometry]{ifsym}

\newcommand{\RevA}[1]{{\color{black}{#1}}}
\newcommand{\RevB}[1]{{\color{black}{#1}}}
\newcommand{\RevC}[1]{{\color{black}{#1}}}
\journalname{Theoretical and Computational Fluid Dynamics}
\begin{document}

\title{Prediction of aerothermal characteristics of a generic hypersonic inlet flow
\thanks{This work was supported by NASA under grant number NNX15AU93A, and AFOSR under grant number FA9550-16-1-0319.}
}
%%\subtitle{WMLES of three-dimensional intersecting shock-wave/turbulent boundary-layer interactions}

\titlerunning{Prediction of aerothermal characteristics of a generic hypersonic inlet flow}        % if too long for running head

\author{Lin Fu $^{\textrm{\Letter}}$ \and
        Sanjeeb Bose   \and
        Parviz Moin
}

\authorrunning{L. Fu et al.} % if too long for running head

\institute{Lin Fu \at
              \emph{Corresponding author.} \\
              Center for Turbulence Research, Stanford University, Stanford CA 94305-3024, USA \\
              \email{linfu@ust.hk}
           \and
           Sanjeeb Bose \at
              Cascade Technologies Inc., Palo Alto, CA 94303, USA
           \and
           Parviz Moin\at
              Center for Turbulence Research, Stanford University, Stanford CA 94305-3024, USA
}

\date{Received: date / Accepted: date}

\maketitle

\begin{abstract}
Accurate prediction of aerothermal surface loading is of paramount importance for the design of high speed flight vehicles.  In this work, we consider the numerical solution of hypersonic flow over a double-finned geometry, representative of the inlet of an air-breathing flight vehicle, characterized by three-dimensional intersecting shock-wave/turbulent boundary-layer interaction at Mach 8.3.  High Reynolds numbers ($Re_L \approx 11.6 \times 10^6$ based on free-stream conditions) and the presence of cold walls ($T_w/T_\circ \approx 0.26$) leading to large near-wall temperature gradients necessitate the use of wall-modeled large-eddy simulation (WMLES) in order to make calculations computationally tractable.
The comparison of the WMLES results with experimental measurements shows good agreement in the time-averaged surface heat flux and wall pressure distributions, and the WMLES predictions show reduced errors with respect to the experimental measurements than prior RANS calculations.  The favorable
comparisons are obtained using a standard LES wall model based on equilibrium boundary layer approximations despite the presence of numerous non-equilibrium
conditions including three dimensionality in the mean, shock-boundary layer interactions, and flow separation.  We demonstrate
that the use of semi-local eddy viscosity scaling (in lieu of the commonly used van Driest scaling) in the LES wall model is necessary to accurately predict the surface pressure loading and heat fluxes.

\end{abstract}

\section{Introduction}
%

%For both military and civil applications, the vehicles flying at highly supersonic or hypersonic speed are promising candidates for future transportation \cite{Leyva2017}. One principal challenge for designing such vehicles is the development of the thermal protection system \cite{candler2019rate}, which can withstand the aerodynamic heating of the surface. The development of efficient and accurate simulation tools that can predict these aerodynamic characteristics (e.g., laminar-turbulent transition, flow separation, heat transfer) can complement experimental campaigns \cite{willems2015experiments}\cite{sandham2014transitional}\cite{schuelein2014effects}\cite{schneider2008development}.  The use of simulation strategies would help reduce trial and error iterations expediting the design process \cite{gaitonde2015progress} and provide additional data that can be otherwise difficult to probe or measure in experimental facilities.

Hypersonic wall-bounded flows for realistic flight vehicles can be characterized by high Reynolds numbers and cold
surface temperatures compared to the free-stream stagnation temperature. The prohibitive computational cost associated
with high Reynolds numbers are well known \cite{choi2012grid}, while the cold wall conditions exacerbate near-wall
resolution requirements associated with the large temperature gradients in the vicinity of peak viscous dissipation.
As a result, direct numerical simulations of these flows have been largely limited to simple geometries and low
Reynolds numbers such as high-speed compressible boundary layer flows \cite{duan2011directMach}, hypersonic boundary-layer transitional flow for a flared cone \cite{hader2019direct}, turbulent boundary layer along a compression ramp \cite{adams2000direct}, and transitional shock/boundary-layer interaction \cite{sandham2014transitional}\cite{Fu2020Direct}.

When more realistic geometries and conditions are considered, the RANS approach is commonly used in industrial settings due to reduced computational costs compared to DNS strategies.  However, RANS based approaches have been demonstrated to have limited accuracy in hypersonic flow regimes; significant errors in peak aerodynamic heating
($\approx 25\%$) \cite{currao2019hypersonic} are observed and macroscopic flow
characteristics are misrepresented, in particular when laminar/turbulent transition or boundary layer separation are present \cite{georgiadis2014status}\cite{fu2013rans}. Cold wall conditions in hypersonic flow regimes also challenge traditional RANS models
(e.g., Spalart-Allmaras, $k-\omega$) to predict near-wall turbulent fluctuations or transverse heat fluxes \cite{huang2019assessment} even in zero-pressure gradient boundary layers.  Algebraic RANS closures, such as the Baldwin-Lomax model \cite{baldwin1978thin},
have been shown to offer reasonable predictions in high Mach number boundary layer
flows \cite{rumsey2010compressibility}.  However, application of these models to the
double-finned inlet flow presently considered show substantial errors in the
surface heat fluxes and in the extent of flow separation
\cite{gaitonde1993calculations}.

%% ($\approx 20\%$)

%%%Compared to the experiment of hypersonic transitional shockwave/boundary-layer interaction with high shock incidence angle, Currao et al. \cite{currao2019hypersonic} show that the RANS model underestimates the peak aerodynamic heating by a factor of up to $25\%$. Fu and Wang \cite{fu2013rans} introduce the correlation based empirical transition model to improve the performance of RANS turbulence models for hypersonic flows, and the universality and generality still remain an open question for realistic engineering flows.

Large-eddy simulations have been shown to offer superior accuracy in the prediction of many of these flow regimes. However, it is well known that near-wall resolution requirements for LES are prohibitive in high Reynolds number conditions.  Alternatively, the WMLES approach, where flow structures that scale with the boundary layer thickness are resolved while effects of unresolved near-wall eddies (at viscous length scales, $l_{\nu} = \nu/u_{\tau}$)
are modeled, has been shown to be computationally tractable for high Reynolds number flows and predictive in several complex flows \cite{bose2018wall}.  In the context of high-speed flows, WMLES has been successfully applied to the
prediction of shock-induced separation \cite{kawai2013dynamic}\cite{souverein2010effect}, oblique shock wave interaction with lateral confinement and boundary layer separation \cite{bermejo2014confinement}\cite{helmer2012three},
transitional flows \cite{mettu2018wall}, and aerodynamic heating \cite{yang2017aerodynamic}\cite{Fu2020Direct}.
However, most of the high speed applications of WMLES have been conducted in
relatively simple geometries or in the absence of technologically  relevant cold wall conditions.

%%%%%%%%%%%%%%%%%%%%%%%%%%%%%%%%%%%%%%%%%%%%%%%%%%%%%%%%%%%%%%%
% discuss iyer and malik later; also consider including references to komives and candler
%Iyer and Malik \cite{iyer2019analysis} investigate the equilibrium wall model for the high-speed turbulent flows by priori and posteriori analyses, and propose an empirical scaling for computing the viscous wall-normal spacing in the damping function of the eddy viscosity model.
%%%%%%%%%%%%%%%%%%%%%%%%%%%%%%%%%%%%%%%%%%%%%%%%%%%%%%%%%%%%%%%

To this end, the present work considers a canonical model of a realistic inlet of an air-breathing hypersonic vehicle \cite{kussoy1992intersecting}. The configuration consists of two sharp fins mounted on a flat plate.  An incident
hypersonic turbulent boundary layer approaches the two vertical fins generating a crossing shock pattern resulting in high local aerothermal loading and flow separation.  The objective of this investigation is to assess the predictive capability of
of wall modeled LES in this complex geometry and flow regime with emphasis on
the prediction of surface heat fluxes, mechanical loading, and separation
that arises from the impinging shock structure. 
\RevB{WMLES results are strictly grid dependent since the grid size $\Delta$ appears in the governing equations through the subgrid and wall model formulae deployed. The models used in the present WMLES do converge to DNS solutions in the limit of very fine resolutions. However, the main question that we seek to answer is whether quantities of interest can be predicted with acceptable accuracy at affordable cost. Specifically, the trade-off between accuracy
and resolution requirements is not well understood in complex flows, particularly those
with mild separations.  This investigation attempts to characterize the threshold 
resolution for sufficiently accurate simulations in this double fin configuration.}

The remainder of this paper is organized as follows. In Section 2, the governing equations, the wall model and the numerical method are briefly reviewed. In Section 3, the flow conditions and corresponding computational setup are described.  Section 4 analyzes the WMLES results, including detailed spatial structures of the mean flow, and assesses their accuracy with respect to existing experimental measurements and prior RANS simulations. It is additionally demonstrated that the cold wall conditions in this configuration necessitate augmentation of wall model eddy viscosity to be scaled on semi-local conditions rather than solely on wall quantities typically used in prior WMLES calculations. Concluding remarks and discussions are provided in Section 5. \RevB{In the Appendix, convergence of the results with longer statistical averaging time (statistical sample size) and finer grid resolution is discussed.}

\section{LES methodology}

\subsection{LES governing conservation equations} %THIS SECTION'S NAME IS MANDATORY

The Favre-averaged compressible Navier-Stokes equations in the conservative form are
\begin{eqnarray}
\label{eq:LES_equation}
\frac{{\partial \overline{\rho} }}{{\partial t}} + \frac{{\partial \overline{\rho} \widetilde{{u}_j}}}{{\partial {x_j}}} &=& 0,\\
\frac{{\partial \overline{\rho} \widetilde{{u_i}}}}{{\partial t}} + \frac{{\partial \overline{\rho} \widetilde{{u_i}}\widetilde{{u_j}}}}{{\partial {x_j}}} + \frac{{\partial \overline{P}}}{{\partial {x_i}}} &=& \frac{{\partial \widetilde{{\sigma _{ij}}}}}{{\partial {x_j}}} - \frac{{\partial {\tau _{ij}^{sgs}}}}{{\partial {x_j}}},\\
\frac{{\partial \overline{E}}}{{\partial t}} + \frac{{\partial [(\overline{E} + \overline{P}){\widetilde{{u_j}}]}}}{{\partial {x_j}}} &=& \frac{{\partial }}{{\partial {x_j}}}(k \frac{{\partial \overline {T}}}{{\partial {x_j}}}) + \frac{{\partial (\widetilde{u_i} {\widetilde{{\sigma _{ij}}})}}}{{\partial {x_j}}} - \frac{{\partial {Q_j^{sgs}}}}{{\partial {x_j}}} - \frac{{\partial \left({\widetilde{u_i}} {\tau _{ij}^{sgs}}\right)}}{{\partial {x_j}}},
\end{eqnarray}
where $\rho$, $P$, and $T$ denote the fluid density, pressure, and temperature, respectively. $u_i$ denotes the velocity component in the $x_i$ coordinate direction. $\overline{E} = \bar{\rho}\tilde{e} + \overline{\rho}\widetilde{{u_k}}\widetilde{{u_k}}/2$ denotes the total energy, $\widetilde{{\sigma _{ij}}} = \mu(\tilde{T}) (\widetilde{S}_{ij}- 1/3 \widetilde{S}_{kk} \delta_{ij})$ is the resolved deviatoric stress tensor, and $\widetilde{S}_{ij} = 1/2({\partial {\widetilde u_i}}/{\partial {x_j}} + {\partial \widetilde {u_j}}/{\partial {x_i}})$ is the resolved strain rate tensor.
The subgrid stress ${\tau _{ij}^{sgs}}$ and heat flux ${Q_j^{sgs}}$ arising from the
effect of unresolved eddies are defined as
\begin{equation}
{\tau _{ij}^{sgs}} = \overline{\rho} (\widetilde {u_i u_j} - \widetilde{u_i} \widetilde{u_j}), \text{ } {Q_j^{sgs}} =  \overline{\rho} (\widetilde {e u_j} - \widetilde{e} \widetilde{u_j}),
\end{equation}
%
%%${\tau _{ij}^{sgs}} - 1/3{\tau _{kk}^{sgs}}\delta_{ij} = -2C \overline{\rho} {\triangle}^2|\widetilde{S}_{ij}| (\widetilde{S}_{ij}- 1/3 \widetilde{S}_{kk} \delta_{ij})$
The subgrid stresses and heat fluxes are closed with the Vreman eddy
viscosity model \cite{vreman2004eddy} supplemented with a constant turbulent
Prandtl number ($Pr_t = 0.9$).  The equation of state for the fluid is a calorically perfect gas, $\bar{P} = \bar{\rho} R \tilde{T}$,
where $R$ denotes the specific gas constant.  The relation between the dynamic viscosity and the temperature is characterized by the Sutherland's law with a model constant $S/T_r = 1.38, T_r = 80$~K, and the Prandtl number is $0.72$. The calorically perfect gas assumption is adopted based on the low free-stream temperature for the configuration under consideration, $T_\infty = 80$ K.
\RevC{Hereafter, the operator symbols $\bar{\cdot}$ and $\tilde{\cdot}$ denote the Reynolds and Favre averages, respectively.  $f^{'}=f-\bar{f}$ denotes the fluctuation defined based on the Reynolds average, and $f^{''}=f-\overline{\rho f}/\bar{\rho}$ denotes the fluctuation defined based on the Favre average.}

%=========================================================================
\subsection{LES wall model based on equilibrium boundary layer approximations}

As the near-wall eddies with length scales characterized by viscous
length scales are not resolved in the present formulation,
their aggregate effect on the wall stress and heat flux must be modeled. (For a
detailed description and review of the wall models for LES, see \cite{bose2018wall}\cite{larsson2016large}\cite{wangmoin2002}).  The present LES wall model assumes that the
pressure gradient and convection effects can be neglected for unresolved
eddies between the wall and the local LES resolution (of grid size, $\Delta$),
and that these eddies reach a statistically stationary state over the duration
of the simulation time step ($\Delta t$).  With these approximations, the
simplified momentum and total energy equations can be \RevB{written as \cite{kawai2012wall}}
\RevC{
\begin{equation}
\label{eq:Equilibrium_wm_1}
\frac{d}{{dy}}\left[ {(\bar{\mu}  + {\bar{\mu} _{t,wm}})\frac{{d{\bar{u}_\parallel }}}{{dy}}} \right] = 0 ,
\end{equation}
\begin{equation}
\label{eq:Equilibrium_wm_2}
\frac{d}{{dy}}\left[ {(\bar{\mu}  + {\bar{\mu} _{t,wm}}){\bar{u}_\parallel }\frac{{d{\bar{u}_\parallel }}}{{dy}} + {c_p}\left(\frac{\bar{\mu} }{{Pr }} + \frac{{{\bar{\mu} _{t,wm}}}}{{{{Pr }_{t,wm}}}}\right)\frac{{d\bar{T}}}{{dy}}} \right] = 0 ,
\end{equation}
where $y$ and $\bar{u}_\parallel$ denote the wall-normal coordinate and the velocity component parallel to the wall, respectively. $c_p$, and $Pr$ denote the specific heat capacity at constant pressure, and the molecular Prandtl number, respectively. Turbulent stresses
and heat fluxes are modeled with an eddy viscosity, $\bar{\mu} _{t,wm}$, given by the
mixing length model,

\begin{equation}
{\bar{\mu} _{t,wm}} = \kappa \bar{\rho} y\sqrt {\frac{{{\bar{\tau} _w}}}{\bar{\rho} }} D ,
\end{equation}
where $\kappa = 0.41$ is the von K\'arm\'an constant,  the damping function $D$ is defined as
\begin{equation}
\label{damping_function}
D = \left[1 - \exp \left( - \frac{{y_{vD}^ + }}{A^ + }\right)\right]^2 ,
\end{equation}
where $A^ + = 17$, and  ${y_{vD}^ + } = {y}/({{{\bar{\nu} _w}/{\bar{u}_\tau }}})$, $\bar{\nu} _w$ and $\bar{u}_\tau$ denote the kinematic viscosity and friction velocity at the wall. However, it is shown in \cite{iyer2019analysis}\cite{yang2018semi} that the van Driest transformation performs poorly in collapsing the compressible velocity profile onto the incompressible counterpart for wall-bounded flows with cold wall condition.}
Due to this, we additionally consider in this work \RevB{a semi-local scaling \cite{huang1995compressible,patel2015semi,iyer2019analysis,muto2019equilibrium}}
\begin{equation}
\label{SL_scaling}
y_{SL}^ +  = \frac{{ \bar{\rho}(y) {\sqrt{\left( {{\bar{\tau} _w}/ \bar{\rho}(y)  } \right)}}y}}{ \bar{\mu}(y) },
\end{equation}

\noindent
where the dynamic viscosity is computed based on the local conditions at a off-wall distance, $y$, and $y_{SL}^+$ is used in place of $y_{vD}^+$ in Eq.~\eqref{damping_function}.  \RevB{
While the semi-local scaling has been used to treat variable property effects close to the 
wall and explore the collapse of canonical equilibrium boundary layers, its \emph{a posteriori} impact for wall modeled LES in a non-equilibrium, hypersonic flow is presently 
not well understood.  This semi-local scaling is shown below to have significant effects 
on the prediction of the non-equilibrium flow, especially in the non-equilibrium boundary 
layer flow between the fins.}

\RevC{The boundary conditions for Eqns.~\eqref{eq:Equilibrium_wm_1} and \eqref{eq:Equilibrium_wm_2} for the velocity and temperature are no-slip, isothermal
conditions at the wall ($\bar{u}_{||}(y=0) = 0, \bar{T}(y=0) = T_w$) and the interior
LES conditions ($\bar{u}_{||}(h_{wm}) = \tilde{u}_{\rm{LES}}, \bar{T}(h_{wm}) = \tilde{T}_{\rm{LES}}$) at a distance, $h_{wm}$,
from the wall.}  In this work, the matching location is chosen as the first off-wall cell center in the local wall-normal direction. 
%Although Kawai and Larsson \cite{kawai2012wall} propose that the matching location should be several cells away from the wall to provide better resolved solution for the wall model as the boundary condition, recent investigations \cite{Fu2018Equilibrium}\cite{lozano2020} have shown that the first-cell coupling is sufficient for the prediction of relevant quantities of interest such as the skin friction or wall heat fluxes.
%This choice is further validated bythis study through the comparisons with the available experimental data in later sections.  
It is important to note that while the wall model does not explicitly contain the non-equilbirium pressure gradient or convection effects, the
influence of these phenomena are implicitly present in the time dependent far field
boundary condition that the interior LES provides at the interface of the wall model region. 

\subsection{Numerical methods}

The compressible, finite-volume code $\rm charLES$ \cite{bres2018large}, is used to conduct the numerical simulations herein. The numerical method consists of a low-dissipation, approximate entropy-preserving scheme and utilizes artificial bulk viscosity to capture solution discontinuities \cite{mani2009suitability}. The LES governing equations are temporally integrated by the explicit third-order strong-stability-preserving (SSP) Runge-Kutta method \cite{gottlieb2001}. The spatial and temporal schemes converge to second- and third-order with respect to the nominal mesh spacing and time step, respectively. Computational meshes based on arbitrary polyhedra are constructed from the computation of Voronoi diagrams 
%associated with the specification of the location of the degrees of freedom clipped against the domain boundaries
\cite{aurenhammer1991voronoi}. 
Details of the numerical method and solver validation campaigns can be found in \cite{Fu2020Direct}, \cite{lozano2020}, \cite{lakebrink2019}, \cite{bres2019modelling}, \cite{bres2018large}, and \cite{lehmkuhl2018}.

\section{Double-finned problem definition and computational setup}

\begin{figure}[!h]
    \centering
    \includegraphics[width=0.6\textwidth]{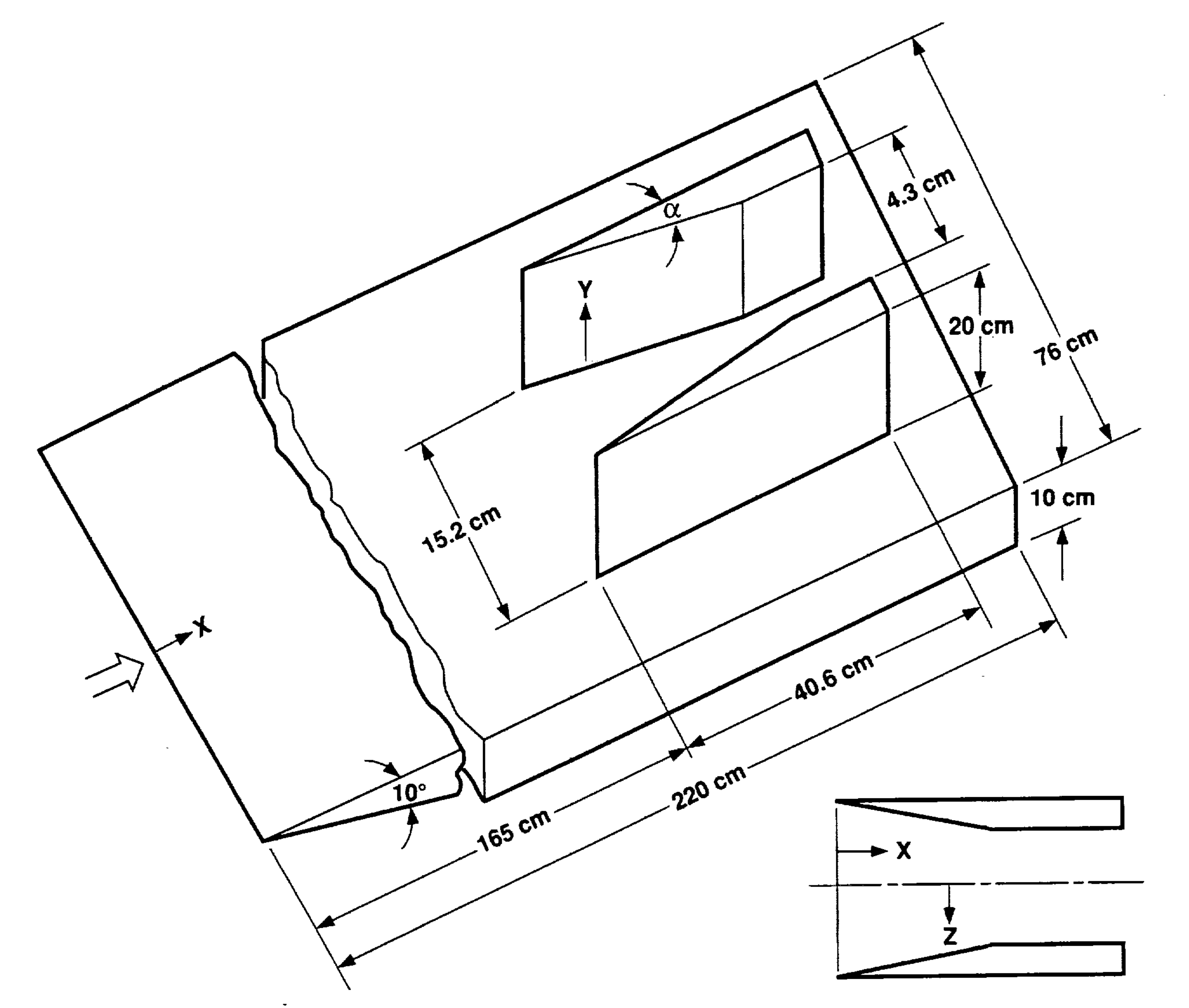}
    \caption{The geometry parameters and the computational coordinate system. The wedge angle for the double fins is $\alpha = 15^\circ$. The arrow indicates that the fluid flows from the leading edge of the flat plate towards the double fins. The leading edges of both the flat plate and the double fins are sharp and the model is symmetric with respect to the plane $z=0$ cm. The free-stream condition is defined $3$ cm ahead of the double fins at $x_\circ = 162$ cm. This figure is adapted from Fig.~1 of \cite{kussoy1992intersecting}.}
    \label{fig:double_fins_parameter}
\end{figure}

The present geometry and computational setup follow those described in the experiment of a $15^\circ$ double-finned configuration \cite{kussoy1992intersecting}. The geometry is composed of two sharp fins with wedge angle $\alpha = 15^\circ$ fastened to a flat plate, as shown in Fig.~\ref{fig:double_fins_parameter}. Specifically, each fin is $20$~cm high and $40.6$ cm long, and the flat plate is $220$~cm long and $10$~cm high. The double fins are placed $165$~cm downstream of the leading edge of the flat plate such that there is sufficient length for a turbulent boundary layer to develop. The free-stream flow measured $3$ cm ahead of the double fins (i.e. at $x_\circ = 162$ cm) has a Mach number $Ma_{\infty} = 8.23$ and Reynolds number $Re_{\delta_\circ} = 1.7 \times 10^5$ based on the local boundary layer thickness ($\delta_\circ = 3.25$~cm). The wall is isothermal
at $T_w = 300$~K which is substantially colder than the stagnation temperature $T_\circ = 1177$ K. In the following discussion, velocity, temperature, density and length are normalized by the sound speed $c_r = 179$~m/s, the reference temperature $T_r= 80$~K, the reference density $\rho_r = 0.0186$~$\rm {kg/m^3}$ and the reference length $L_r = 1$~cm.

\begin{figure}[!h]
    \centering
    \includegraphics[width=0.8\textwidth]{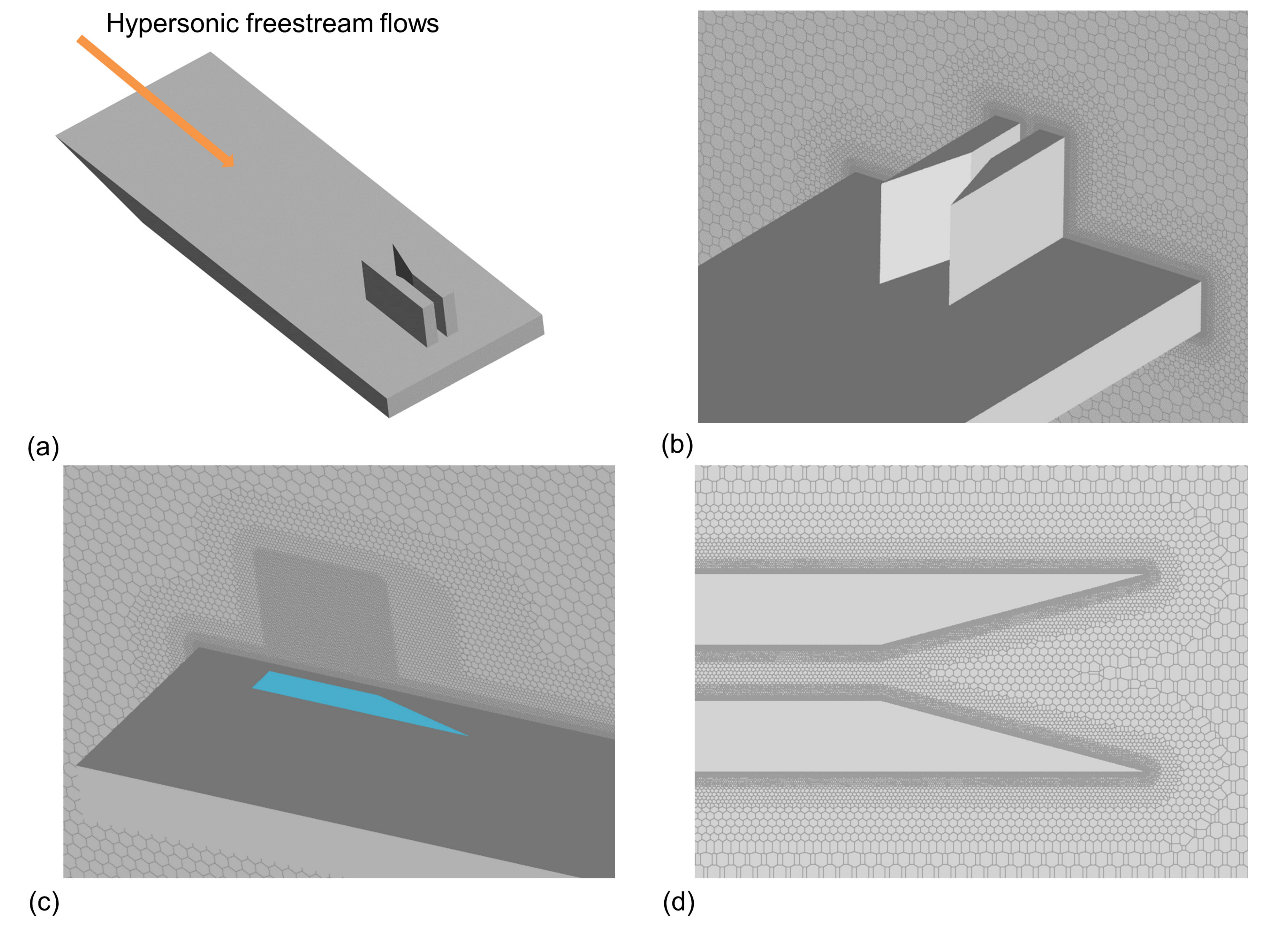}
    \caption{Computational geometry and mesh sketch for the double-fin simulations: (a) the overview of the computational geometry; (b,c,d) the zoom-in views of the Voronoi mesh distributions (70M total mesh elements).}
    \label{fig:double_fins}
\end{figure}

The computational geometry is given in Fig.~\ref{fig:double_fins}(a). The computational domain is sufficiently large to minimize artificial reflections from the far-field outflow boundaries. 

%\RevA{Due to the lack of turbulence statistics from the experiment for the boundary layer between the flat plate leading edge and the double-fin entrance, fully turbulence assumption is adopted without considering the potential transition process following \cite{gaitonde1995structure}\cite{narayanswami1993numerical}}. 
\RevA{The inlet flow condition is imposed by combining a uniform flow with turbulence fluctuations generated by a synthetic turbulence generation method \cite{wu2017inflow}.  The entry length of the domain is not of sufficient extent to replicate the experimental profiles at the inlet. However, the flow between the fins is expected to be insensitive to the details of the incoming boundary layer owing to significant geometrical and physical effects present.  Freestream conditions upstream of the sharp leading edges are adjusted such that
the Mach number (behind the leading edge shock) matches the experimental
measurements upstream of the double fin entrance.}
%The turbulence kinetic energy is set as about $2\%$ of that of the mean flow. The rationale for this higher Mach number is that the inlet flow will be decelerated by the oblique shockwave induced by the sharp leading edge of the flat plate and the present setup leads to a good match of flow status with the experimental free-stream flow at $3$~cm ahead of the double fins.
The computational domain is discretized with the $7 \times 10^7$ Voronoi mesh elements adaptively clustering near the wall, as shown in Fig.~\ref{fig:double_fins}(b,c,d). Based on the resolution of the finest Voronoi mesh element near the wall, the turbulent boundary layer at $x_\circ = 162$ cm is resolved by approximately 40 cells. (The present resolution is much coarser than that previously employed for studying the confinement effects in shock wave/turbulent boundary layer interactions, see Table 1 of \cite{bermejo2014confinement}). The resolution is coarsened further away from the wall to a maximum of $\approx 0.1\delta_\circ$ (see Fig.~\ref{fig:double_fins}(d)).
%It is infeasible to further coarsen the deployed grid for the following two reasons. First, further coarsening the grid will lead to insufficient resolution to properly capture the boundary layer growth from the sharp leading edge. Second, the secondary separation on the plate is small, and this resolution is sufficient to capture the secondary separation bubble (see Fig.~\ref{fig:Center_plane_streamlines}).  Capturing the flow separation is necessary in order to properly predict the surface heat flux distribution.  Nonetheless, the present first off-wall mesh resolution is roughly 400-time coarser when compared to that required by the RANS simulations (see Table 2 of \cite{narayanswami1993numerical}).

%
\begin{table}[h!]
  \begin{center}
    \caption{\RevC{Grid parameters inside the turbulent boundary layer upstream of the double fins.}}
    \label{tab:double_fins_grid}
    \begin{tabular}{c|c}
      $(\Delta x, \Delta y, \Delta z)/\delta_\circ \mid_{min}$ & $(\Delta x, \Delta y, \Delta z)/\delta_\circ \mid_{max}$ \\
      \hline
      $2.5 \times 10^{-2}$ & $1.0 \times 10^{-1}$ \\
      \hline
    \end{tabular}
        \begin{tablenotes}
        \item[a] \RevC{Note that the minimum and maximum mesh spacings in the plus unit are 9.2 and 36.8, respectively.}
    \end{tablenotes}
  \end{center}
\end{table}
\section{Results and discussions}

In this section, the numerical results from WMLES with the semi-local scaling based damping function (Eq.~(\ref{SL_scaling})) are analyzed
and compared against the experimental measurements. The predictions of the WMLES using the van Driest scaling (Eq.~(\ref{damping_function})) will also be assessed. \RevC{Hereafter, the operator symbol $<{\cdot}>$ denotes the time- and spanwise-average.} \RevB{The main turbulence statistics are collected within a time interval, which is about 11 flow-through times from the fin leading edge to the trailing edge. In the appendix, simulations with additional 8 flow through times averaging interval are presented demonstrating the adequacy of the statistical sample for the main quantities of interest (pressure and heat transfer profiles). }

\subsection{WMLES with semi-local scaling based damping function}
\subsubsection{Overall statistics}

\begin{figure}[!h]
    \centering
    \includegraphics[width=0.8\textwidth]{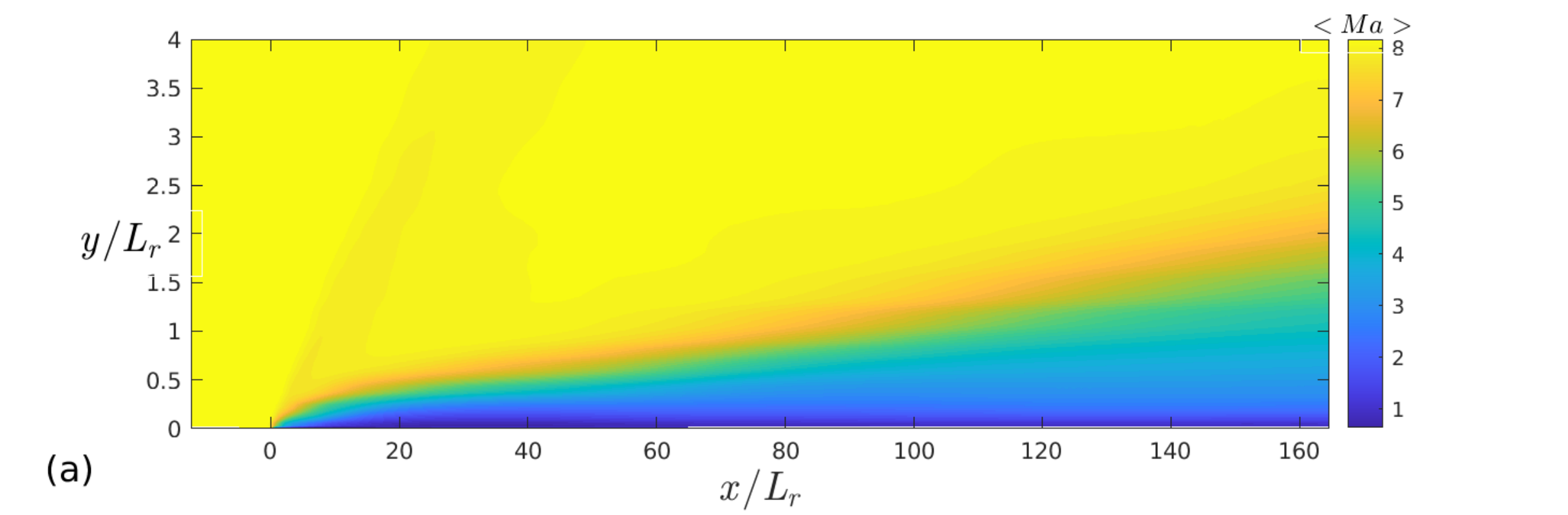}
    \includegraphics[width=0.8\textwidth]{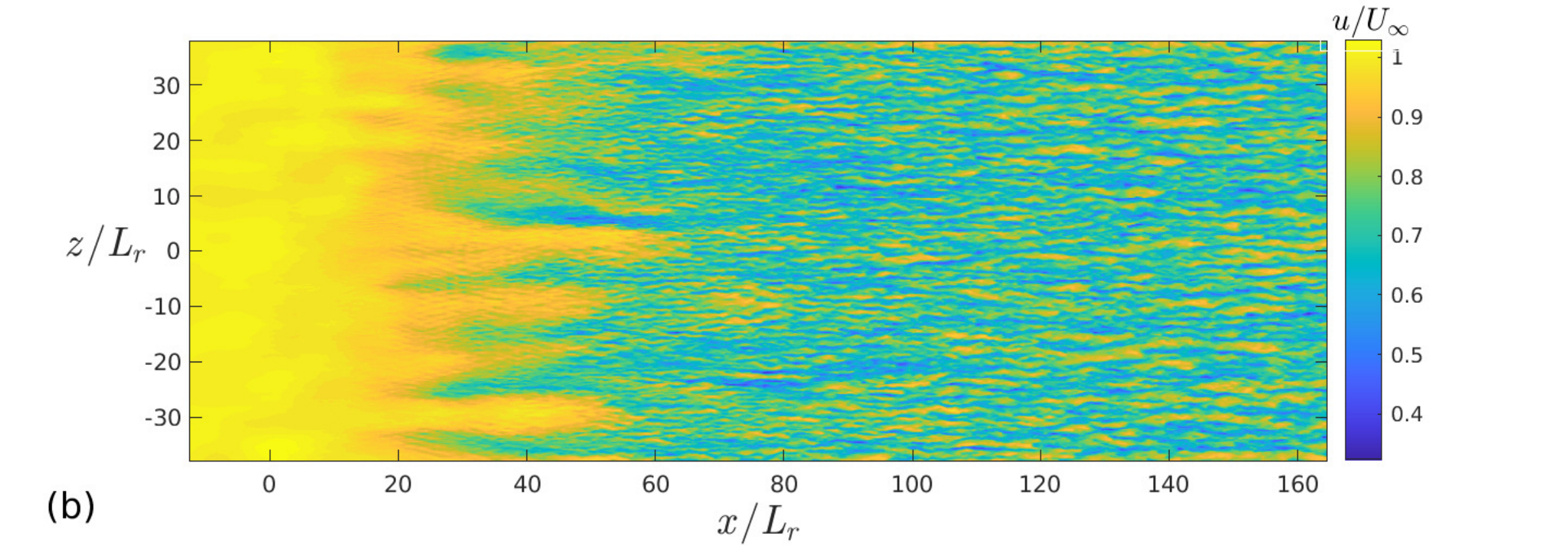}
    \caption{(a) Time- and spanwise-averaged Mach number contour on the wall-normal x-y plane and (b) the instantaneous streamwise velocity distribution on the wall-parallel x-z plane of $y/L_r = 0.3616$. In panel (b), the high-speed region is associated with the oblique shock wave originating from the leading edge of the flat plate at $x/L_r = 0$.}
    \label{fig:flat_plate_bounadry_layers}
\end{figure}

\begin{figure}[!h]
    \centering
    \includegraphics[width=1.0\textwidth]{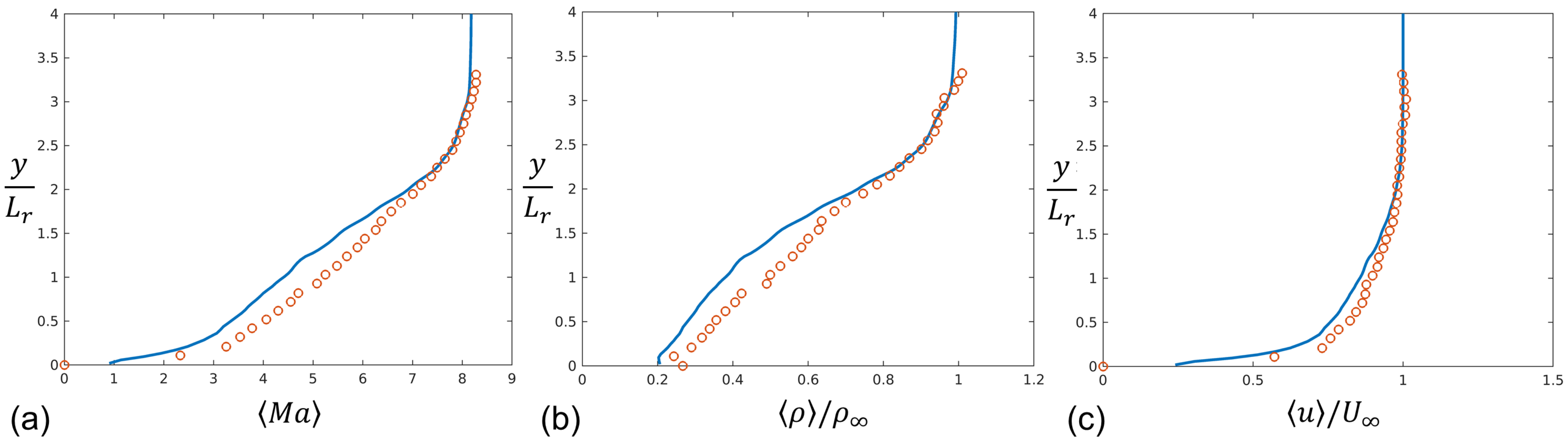}
    \caption{\RevA{Time- and spanwise-averaged distributions of (a) Mach number, (b) density, and (c) streamwise velocity at $x/L_r = 162$. The blue lines and the red circles denote the results from WMLES and experiment \cite{kussoy1992intersecting}, respectively.}}
    \label{fig:double_fins_freestream_compare}
\end{figure}

Fig.~\ref{fig:flat_plate_bounadry_layers} shows the time- and spanwise-averaged Mach number contour on a wall-normal plane and the instantaneous streamwise velocity distribution on a wall-parallel plane at $y/L_r = 0.3616$ between the leading edge of the flat plate and that of the double fins. A weak shockwave is generated at the leading edge of the flat plate, and slightly decreases the Mach number downstream as shown in Fig.~\ref{fig:flat_plate_bounadry_layers}(a). The instantaneous streamwise velocity field in Fig.~\ref{fig:flat_plate_bounadry_layers}(b) shows that the boundary layer transitions, and eventually becomes fully turbulent ahead of the double fins.  The turbulent boundary layer appears sustained for approximately $25\delta_0$ upstream of the \RevA{double finned} entrance. More quantitative comparisons of the flow statistics between the experimental data and the WMLES results at $x/L_r = 162$, just upstream of the fins, are given in Fig.~\ref{fig:double_fins_freestream_compare}. While all the statistics close to the boundary layer edge are in good agreement, there are notable discrepancies inside the boundary layer.  The discrepancies are in part due to the lingering effect of artificial inflow conditions and relatively short developing length of the incoming boundary layer from the sharp leading edge of the plate. \RevA{Since the detailed boundary layer statistics between the flat plate leading edge and the fin entrance are unavailable from the experiment, the pursuit of an exact match of all flow profiles between WMLES and experiment at $x/L_r = 162$ is not realistic. However, as discussed earlier, the flow inside the inlet geometry (between the fins which has been subjected to considerable distortion) may not be as sensitive to the details of the boundary layer flow at the entrance. } 

%It is worth noting that similar discrepancies have been observed in RANS simulations with different turbulence models, see e.g. Table 1 of \cite{narayanswami1993numerical}.}

%

\begin{figure}[!h]
    \centering
    \includegraphics[width=0.5\textwidth]{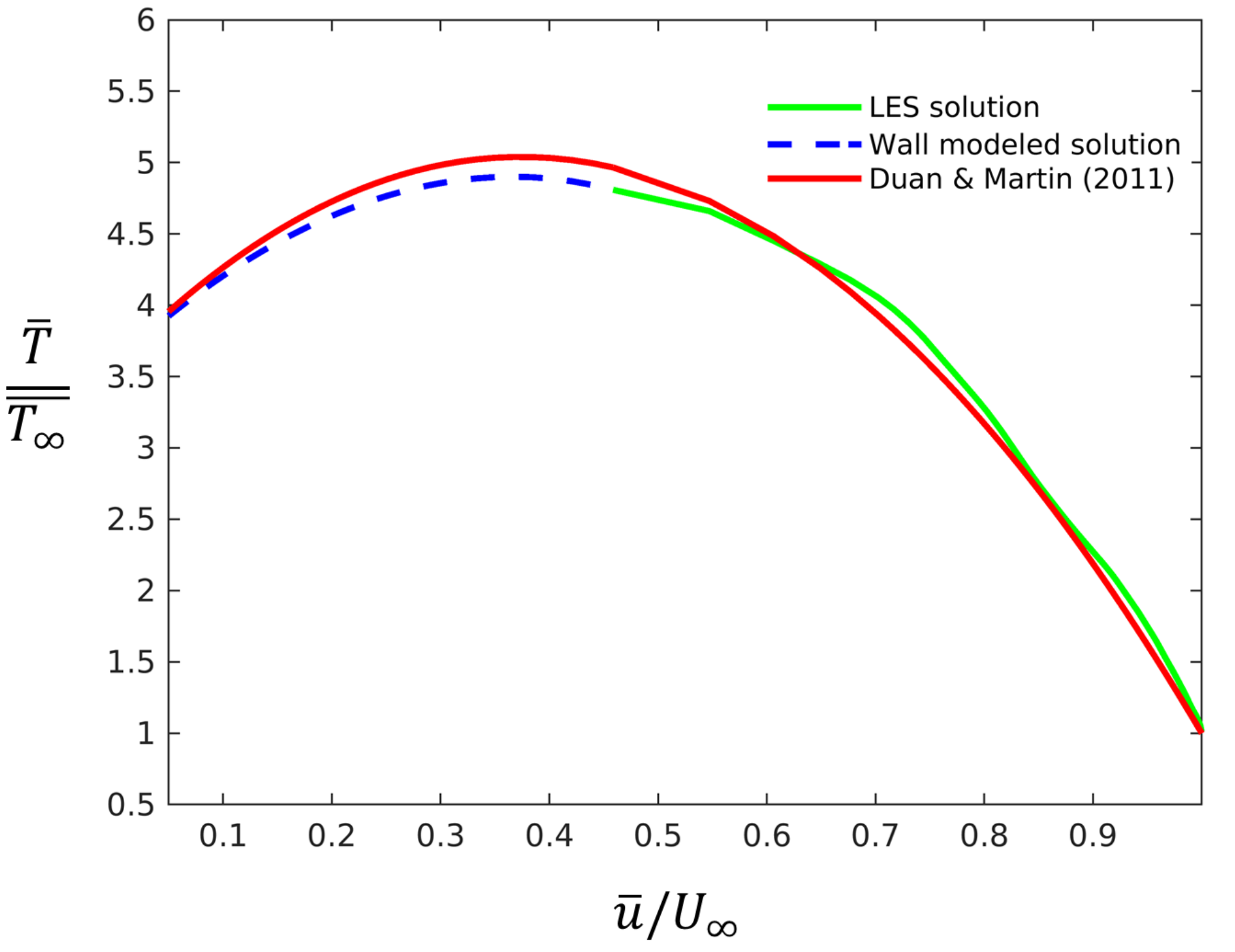}
    \caption{\RevB{Relation between the time-averaged profiles of temperature and velocity at $x/L_r = 162$. Also shown is the model prediction from Duan \& Martin \cite{duan2011direct}.}}
    \label{fig:mean_TV}
\end{figure}

\RevB{The polar plot of the time-averaged profiles of temperature and velocity at $x/L_r = 162$ is shown in Fig.~\ref{fig:mean_TV}. It is observed that the present WMLES solution agrees with the model prediction of Duan \& Martin \cite{duan2011direct} very well.}

\begin{figure}[!h]
    \centering
    \includegraphics[width=0.8\textwidth]{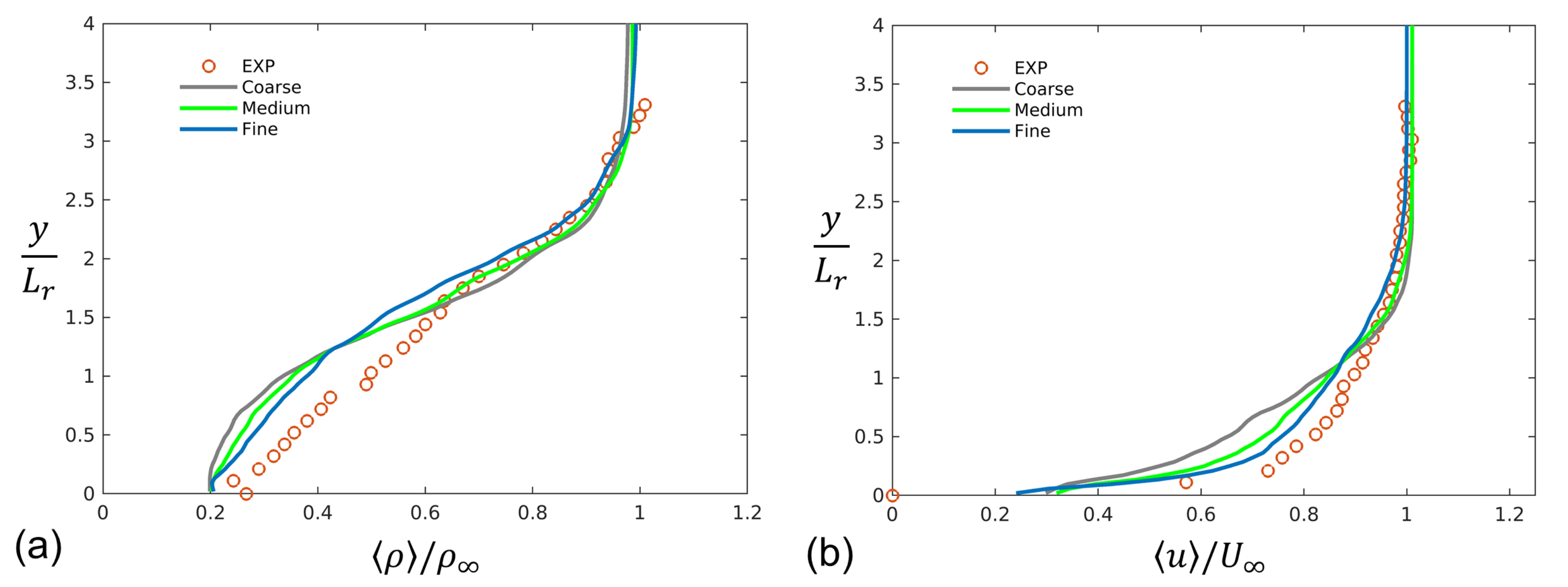}
    \caption{\RevC{Time- and spanwise-averaged distributions of (a) density and (b) streamwise velocity at $x/L_r = 162$. The fine grid denotes the mesh with parameters given in Table~\ref{tab:double_fins_grid}. The resolutions of the medium and coarse grids are $50\%$ and $70\%$ coarser than that of the fine grid in each coordinate direction, respectively. EXP denotes the experimental data \cite{kussoy1992intersecting}.}}
    \label{fig:double_fins_mesh_convergence}
\end{figure}

The results of a grid convergence study at $x/L_r = 162$ is shown in Fig.~\ref{fig:double_fins_mesh_convergence}. The fine grid denotes the mesh with parameters given in Table~\ref{tab:double_fins_grid}. The resolutions of the medium and coarse grids are $50\%$ and $70\%$ coarser than that of the fine grid in each coordinate direction, respectively. The mean streamwise velocity just upstream of the fins exhibits considerable sensitivity to the grid resolution, with profiles from finer grid resolutions moving monotonically closer to the experimental data.  In particular, the boundary layer thickness predicted from the medium and coarse grids is $38\%$ smaller than that given by the experiment, and consequently the local effective Reynolds number differs from the experimental setup as well. With the fine grid, the agreement in terms of the boundary layer thickness is good. The mean density profile is less sensitive to grid resolution. Hereafter, only the simulation results from the fine grid will be discussed and compared with the experimental data at downstream locations.

\begin{figure}[!h]
    \centering
    \includegraphics[width=0.4\textwidth]{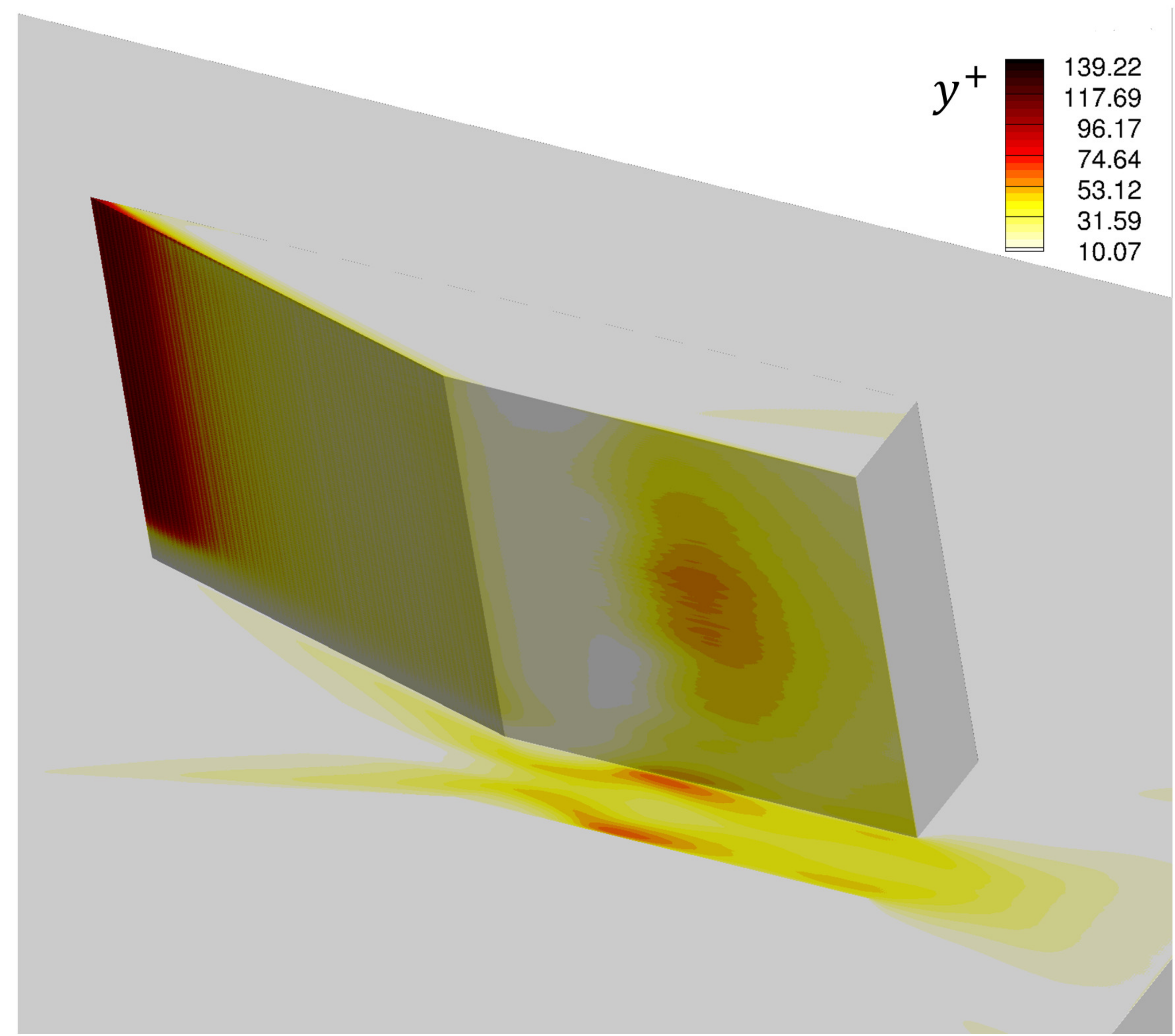}
    \caption{Distribution of the time-averaged $y^+$ at the first off-wall cell centers. For facilitating the presentation, only the data over the flat plate and one vertical fin are shown.}
    \label{fig:double_fins_yplus_avg}
\end{figure}

Fig.~\ref{fig:double_fins_yplus_avg} shows the time-averaged $y^+$ at the first off-wall cell centers, i.e. the matching locations for the wall model. The largest $y^+$ appears around the leading edges of the double fins and the regions where the shock waves impinge on the surfaces. \RevA{It is noticed that, in the regions upstream of the double-fin entrance, the height of the matching location in plus unit is less than $10$. For the downstream regions where shock/boundary layer interaction occurs, the near-wall eddies and the viscous sublayer are not directly resolved in the simulations and the wall model plays a pivotal role in the predicted flow states.}
%The $y^+$ values near the shock impingement location are below $y^+ \approx 20$ suggesting that the details of the LES wall model damping function, $D(y^+)$, may significantly impact the flow solutions.

%

\begin{figure}[!h]
    \centering
    \includegraphics[width=0.9\textwidth]{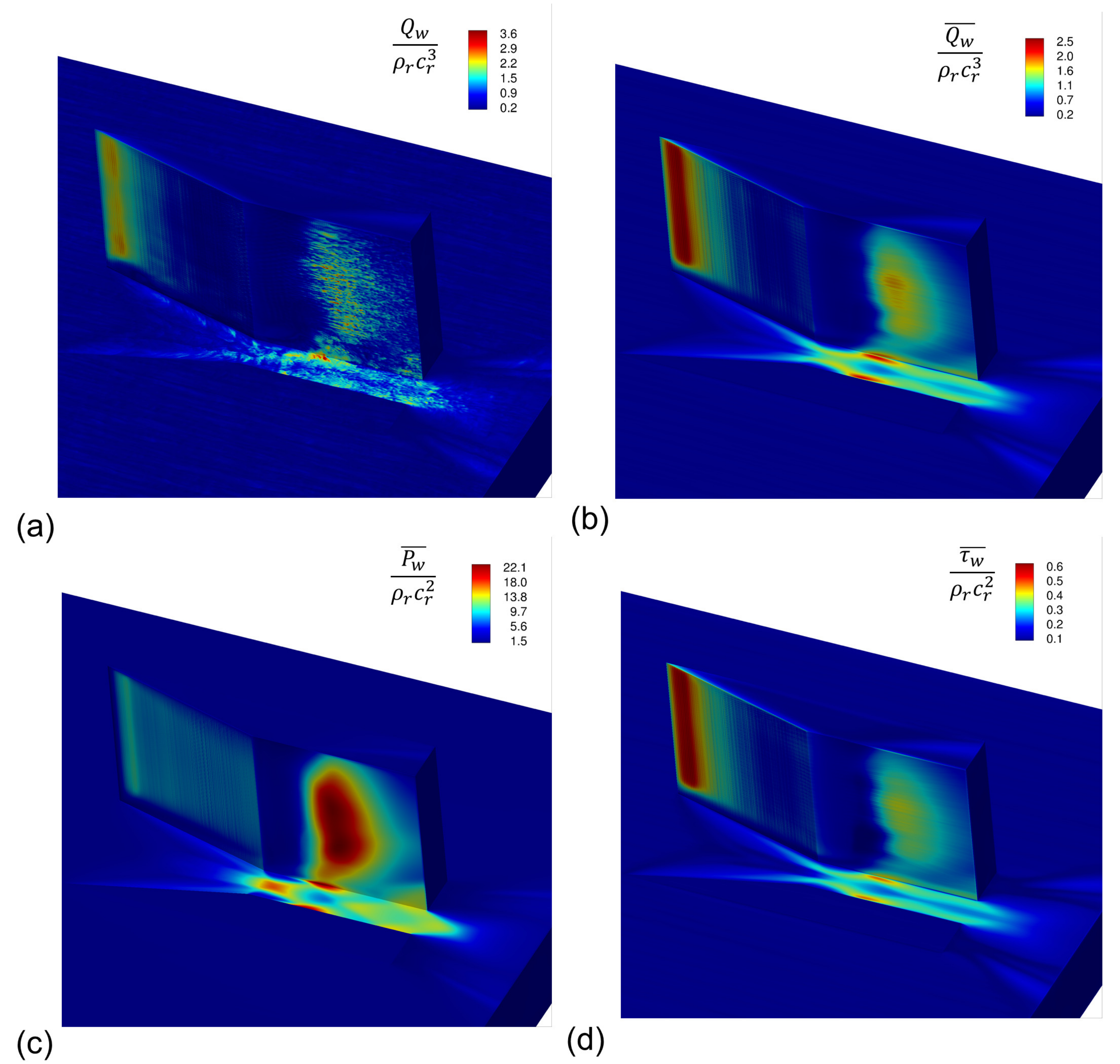}
    \caption{(a) Instantaneous surface heat flux distribution, (b) time-averaged surface heat flux distribution, (c) time-averaged pressure distribution, and (d) time-averaged surface shear stress distribution. The location of the double-shock intersection based on the inviscid theory is around the midpoint of the two fin shoulders. For facilitating the presentation, only the data over the flat plate and one vertical fin are shown.}
    \label{fig:double_fins_flow_field}
\end{figure}

The instantaneous and time-averaged surface heat flux distributions, surface pressure, and surface shear stress distributions are shown in Fig.~\ref{fig:double_fins_flow_field}. As shown in Fig.~\ref{fig:double_fins_flow_field}(a), the instantaneous surface heat flux fluctuates significantly after the double shockwaves induced by the fin leading edges intersect around the shoulders. As shown in Fig.~\ref{fig:double_fins_flow_field}(b)(c)(d), right downstream of the shock waves intersection, the distributions of the time-averaged surface heat flux, surface pressure and surface stress attain local maxima around the centerline of the plate. The peak aerodynamic heating and friction  occur around the shock impingement locations on the fin surfaces.

\subsubsection{Data analyses in x-z and x-y planes}

\begin{figure}[!h]
    \centering
    \includegraphics[width=1.\textwidth]{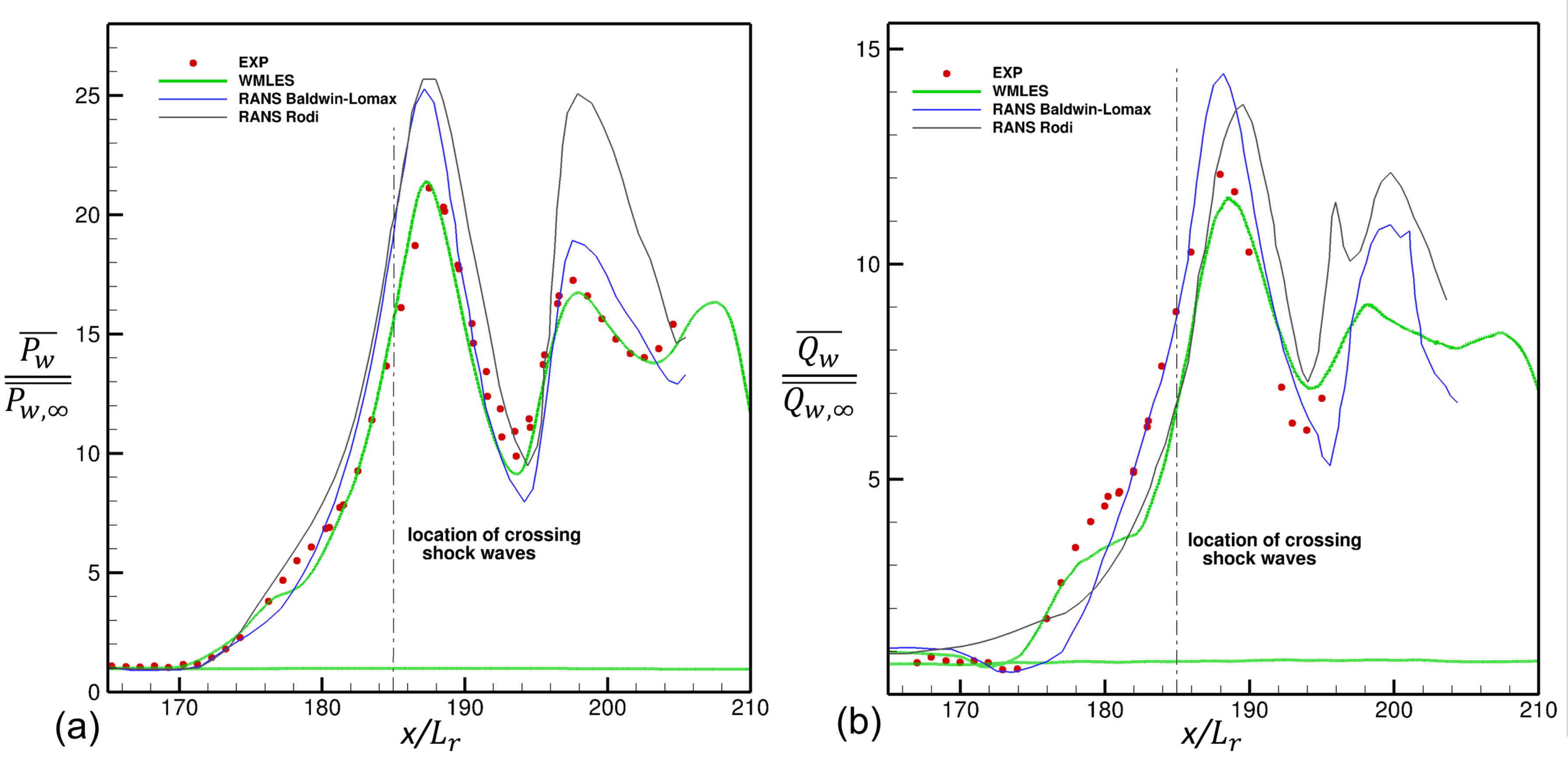}
    \caption{\RevC{Streamwise distributions of the time-averaged (a) surface pressure and (b) surface heat flux on the flat plate at $z/L_r = 0$. The green lines and the red dots denote the results from the WMLES simulation and the experiment, respectively. The location of the double-shock intersection based on the inviscid theory is also shown in the plots.} \RevA{Also shown are the results from the zero-equation Baldwin-Lomax model and the two-equation $k-\epsilon$ model \cite{narayanswami1993numerical}. $\overline{P_{w,\infty}}$ and $\overline{Q_{w,\infty}}$ denote the mean wall pressure and heat flux defined at the location $x/L_r = 162$, where the so-called ``freestream'' condition is provided by the experimental report \cite{kussoy1992intersecting}.}}
    \label{fig:double_fins_pressure_heat}
\end{figure}

\begin{figure}[!h]
    \centering
    \includegraphics[width=0.8\textwidth]{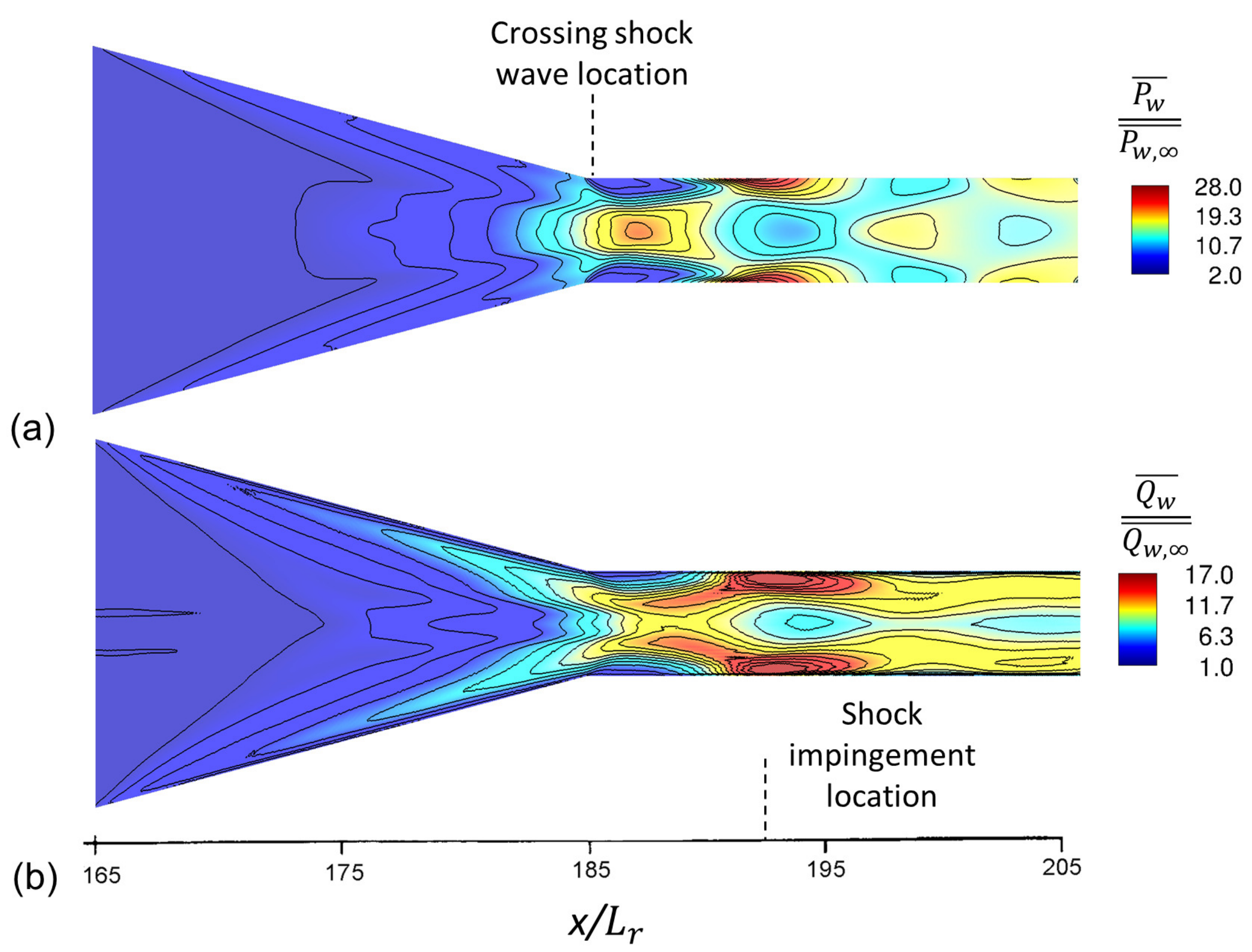}
    \caption{Distributions of the time-averaged (a) surface pressure and (b) surface heat flux on the flat plate at $y/L_r = 0$. The location of the double-shock intersection based on the inviscid theory is also shown in the plots. Also shown is the shock impingement location around $x/L_r = 192$.}
    \label{fig:pressure_heat_plate}
\end{figure}

The time-averaged surface pressure and heat flux distributions along the centerline of the plate between the two fins as well as the double-shock intersection location based on the inviscid theory are given in Fig.~\ref{fig:double_fins_pressure_heat}. The predicted time-averaged pressure distribution from WMLES is in good agreement with the experiment including in the region downstream of the shocks intersection. The static pressure first increases significantly due to the shockwave intersection and subsequently exhibits a rapid drop due to the expansions emanating from the fin shoulders, as depicted by Fig.~\ref{fig:pressure_heat_plate}(a). The peak surface pressure \RevC{$\overline{P_w}/\overline{P_{w,\infty}} \approx 22$} is considerably lower than the prediction from the inviscid theory, \RevC{$\overline{P_w}/\overline{P_{w,\infty}} \approx 45$} \cite{gaitonde1995structure}. Further downstream at $x/L_r \approx 198$, a smaller pressure peak appears due to the second crossing of the reflected shock waves. In terms of the heat flux distribution, the agreement with the experimental data is also favorable across the entire channel between the double fins. The streamwise variation of the surface heat flux follows that of the surface pressure qualitatively. Both the experiment and the WMLES results exhibit an initial decline at $x/L_r \approx 170$, and the predicted heat flux is $20\%$ smaller than that from the experiment in the pre-shock region of $x/L_r \approx 180$, which is the location of a secondary (small) flow separation (see the discussions of Fig.~\ref{fig:skin_flat_plate}). Downstream of the shock wave intersection, the peak heat flux shows $4\%$ discrepancy between the WMLES results and the experimental data. Similar differences are also observed further downstream at $x/L_r = 194$ in the low pressure region and heat flux valley (see also Fig.~\ref{fig:pressure_heat_plate}). Nevertheless, the present prediction of both quantities shows a much better agreement with the experiment than those from the RANS simulations \cite{gaitonde1995structure}\cite{narayanswami1993numerical}. In the RANS solutions, the heat flux plateau around $x/L_r = 180$ upstream of the shock wave intersection is completely missed. 
%and great sensitivity is noticed with regard to the choice of length scale definition in the Baldwin-Lomax model (see Fig.~4 and Fig.~5 in \cite{gaitonde1995structure}).  
Both the zero-equation Baldwin-Lomax model and the two-equation $k-\epsilon$ model overpredict the peak pressure and the peak heat flux by about $20\%$ (see Fig.~3 and Fig.~9 in \cite{narayanswami1993numerical}). As shown in Fig.~\ref{fig:pressure_heat_plate}, the predicted nominal shock impingement location on the side fins is around $x/L_r = 192$ and is similar to the RANS predictions (see Fig.~3 of \cite{gaitonde1995structure}).

\begin{figure}[!h]
    \centering
    \includegraphics[width=0.8\textwidth]{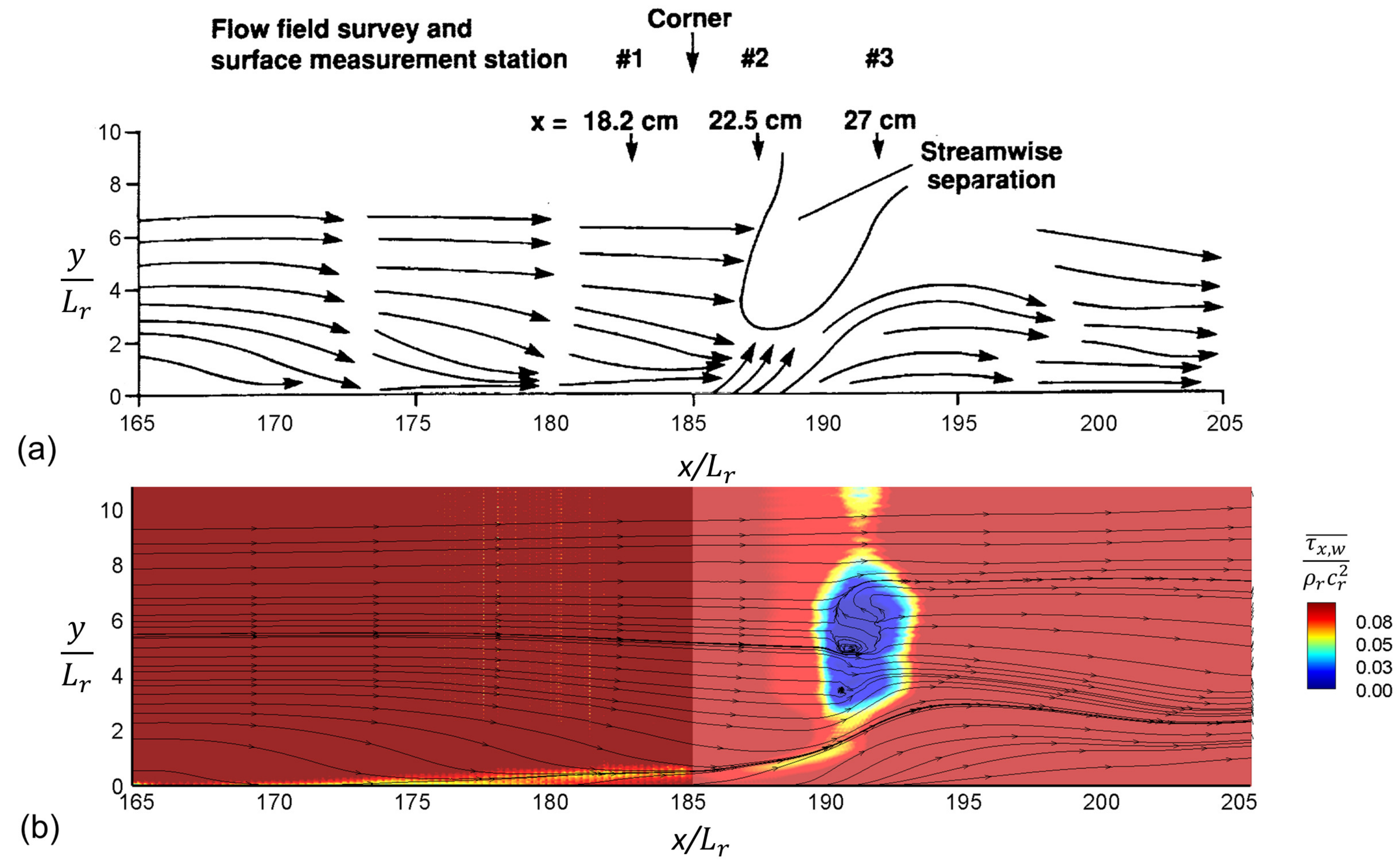}
    \caption{Time-averaged surface skin friction lines on the right fin. Panel (a) denotes the sketch from the experiment and adapted from Fig.~9(b) of \cite{kussoy1992intersecting}. Panel (b) denotes the present WMLES result and the quantity in the contour represents the time-averaged streamwise skin friction.}
    \label{fig:skin_friction_fin}
\end{figure}

\begin{figure}[!h]
    \centering
    \includegraphics[width=0.8\textwidth]{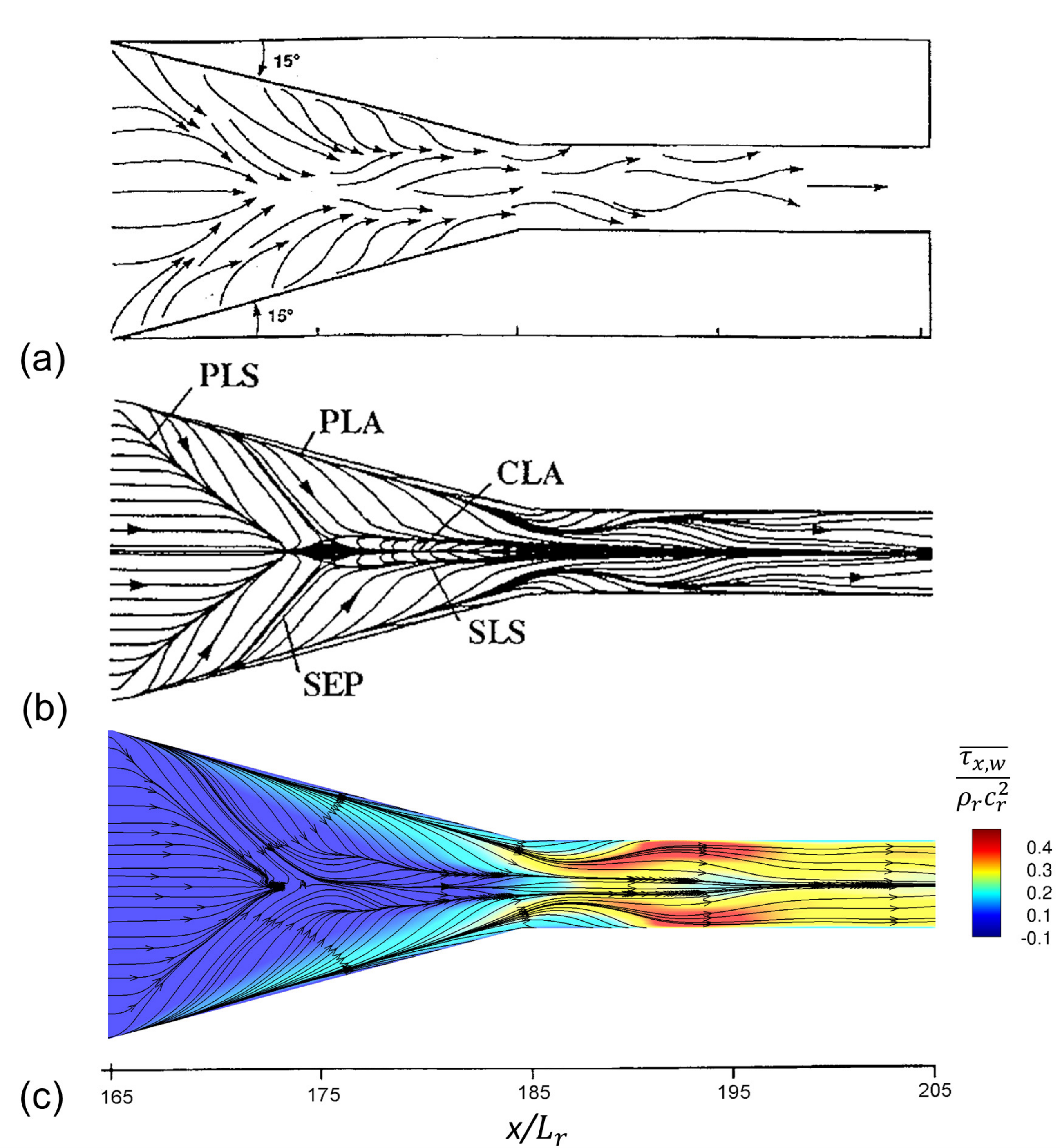}
    \caption{The time-averaged surface skin friction lines on the flat plate. Panel (a) denotes the sketch from the experiment and adapted from Fig.~7(b) of \cite{kussoy1992intersecting}. Panel (b) denotes the CFD result of RANS approach and adapted from Fig.~9 of \cite{gaitonde1995structure}. In Panel (b), PLS denotes the principal line of separation, PLA denotes the principal line of attachment, CLA denotes the center line of attachment, SEP denotes the separatrix, and SLS denotes the secondary line of separation. Panel (c) denotes the present WMLES result and the colored contour levels represent the time-averaged streamwise skin friction.}
    \label{fig:skin_flat_plate}
\end{figure}

To characterize the boundary-layer flow separation, Fig.~\ref{fig:skin_friction_fin} shows the time-averaged surface skin friction lines on the right fin and the corresponding {\it sketch} from the experiment. It is observed that the flow separates around the region where the shock wave impinges on the fin surface. While the overall agreement is good, the predicted separation bubble close to the fin surface starts at $x/L_r=190$, which is delayed compared to the {\it sketch} from the experiment at $x/L_r=187$. As summarized in \cite{narayanswami1993numerical}, the two-equation $k-\epsilon$ model does not capture this separation bubble.

\begin{figure}[!h]
    \centering
    \includegraphics[width=0.7\textwidth]{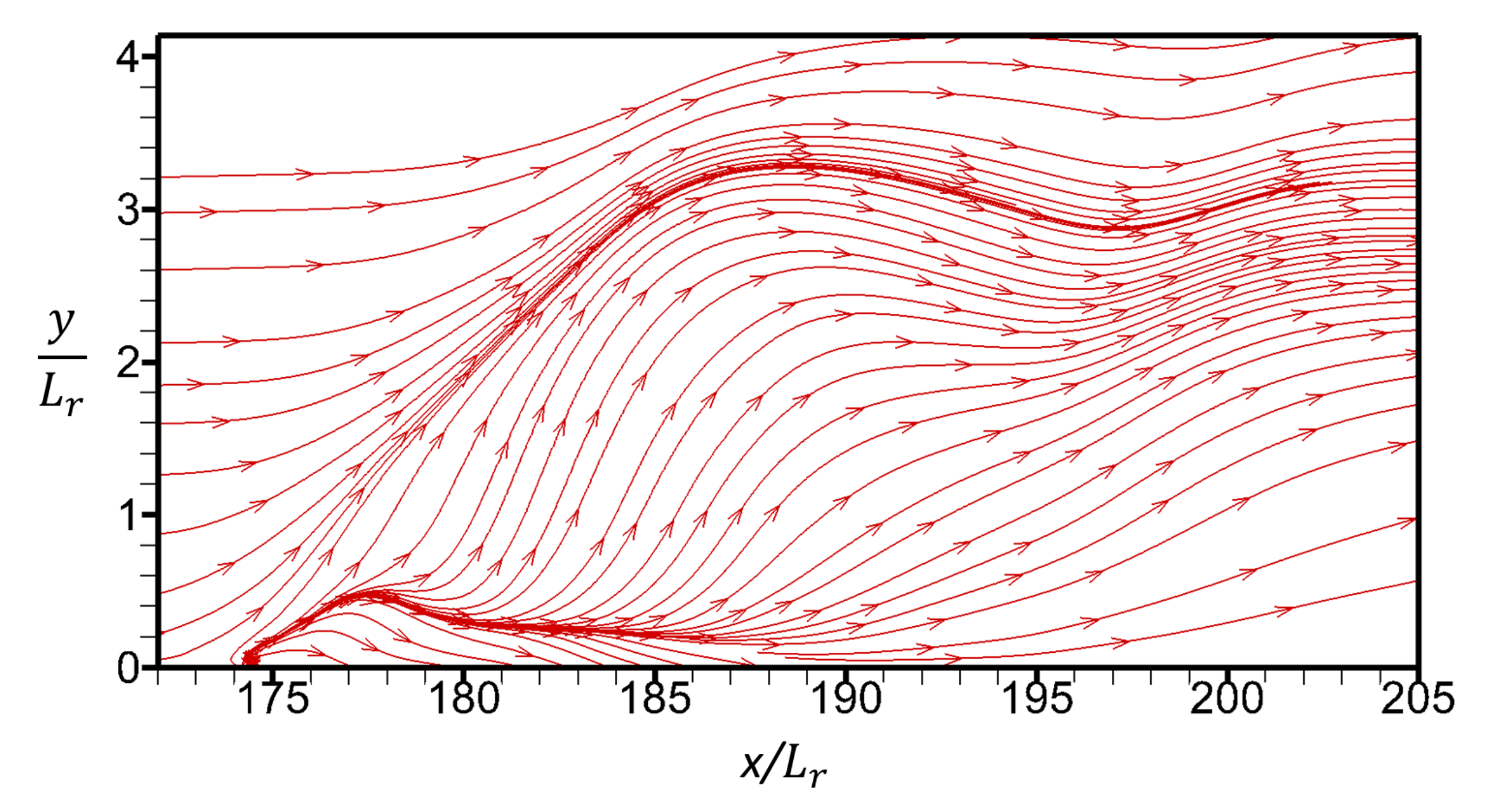}
    \caption{The zoom-in view of the time-averaged streamlines on the symmetry plane. Around $x/L_r \approx 174.5$, there is a critical stagnation point located very close to the plate.}
    \label{fig:Center_plane_streamlines}
\end{figure}

The skin friction lines on the flat plate are given in Fig.~\ref{fig:skin_flat_plate}. The WMLES result is in qualitative agreement with the sketch deduced from experimental measurements, and show similarities with the RANS solution using the Baldwin-Lomax model \cite{gaitonde1995structure}.  There are two lines of coalescence, the principal line of separation (PLS) and the secondary line of separation (SLS). Accordingly, two lines of divergence are also well captured, i.e. the principal line of attachment (PLA) and the center line of attachment (CLA). Close to the centerline, the secondary separation is formed near $x/L_r=175$ and characterized by a pair of streamwise counter-rotating vortical structures. The secondary separation continuously shrinks and disappears near $x/L_r = 197$, where the upside and downside SLS lines converge to the CLA line, i.e. the centerline of the double fins. Fig.~\ref{fig:Center_plane_streamlines} shows the zoom-in view of the time-averaged streamlines on the symmetry plane. The maximum height of the secondary separation is roughly $0.5$ cm at $x/L_r \approx 178$, which is much smaller than the inlet boundary layer thickness of $3.25$ cm around the leading edge of the double fins, and the predicted flow structure is consistent with that reported in Fig.~11 of \cite{gaitonde1995structure}.  
%This secondary separation bubble size is resolved with $O(5)$ points across its height in the present case.

%

\begin{figure}[!h]
    \centering
    \includegraphics[width=0.7\textwidth]{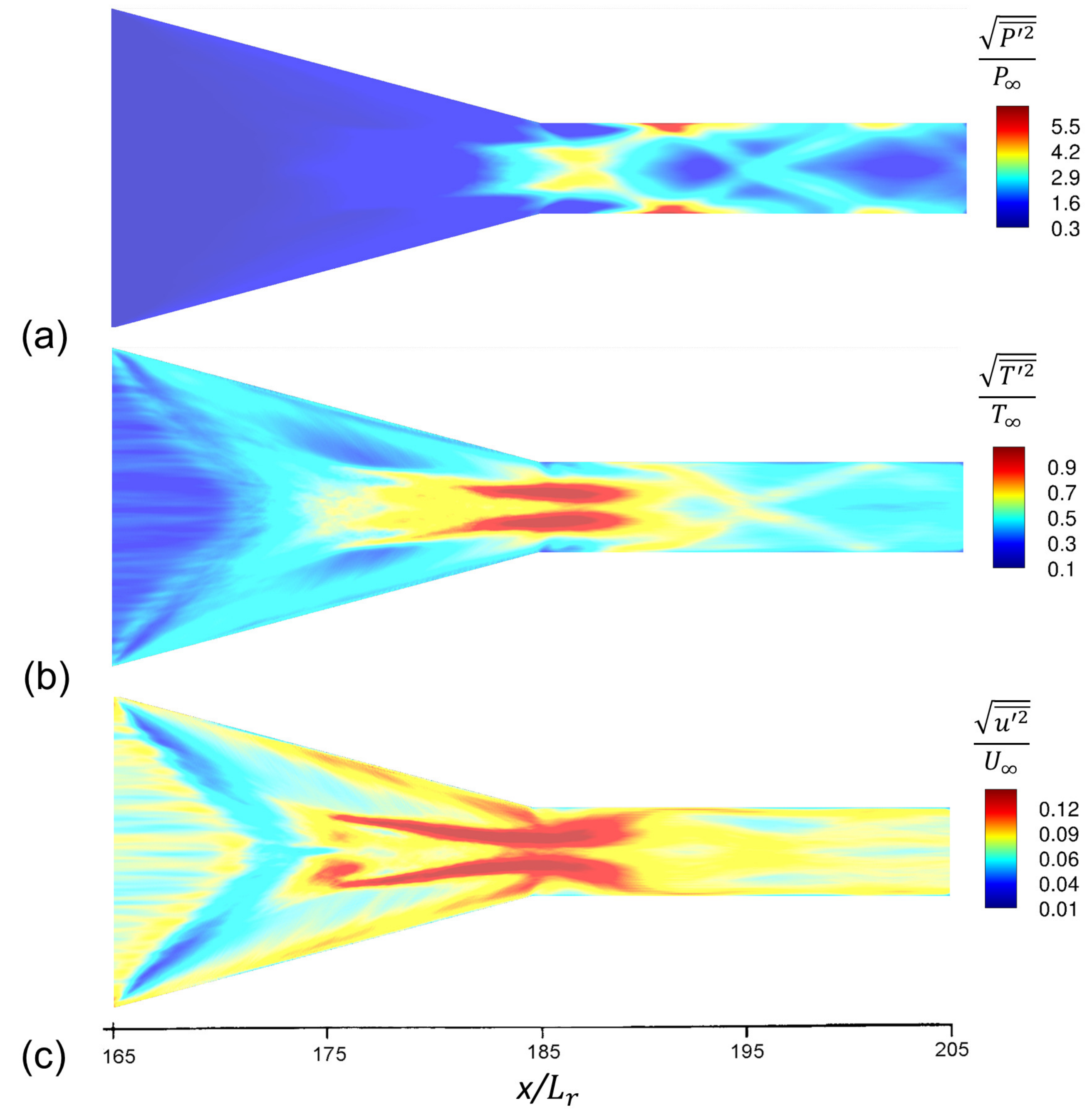}
    \caption{r.m.s statistics of the (a) pressure, (b) temperature and (c) streamwise velocity fluctuations. These statistics are projected to the plate surface from the first off-wall cell centers.}
    \label{fig:RMS_statistics_plate}
\end{figure}

The near-wall root-mean-square (r.m.s) statistics of the pressure, temperature and streamwise velocity are given in Fig.~\ref{fig:RMS_statistics_plate}. The near-wall peak temperature and streamwise velocity fluctuations mainly occur around the secondary lines of separation, in particular at the location where the shockwaves intersect. On the other hand, the peak pressure fluctuations occur on the fin surfaces.

%\RevC{Concerning the boundary-layer development on the side fins, the distributions of the instantaneous streamwise vorticity on the wall-parallel planes at $y/L_r = 2.5375$ and $5.075$ are visualized in Fig.~\ref{fig:Plate_YYY_111_222_vorticity}. The large-scale flow structures induced by the primary vortex can be identified from the streamwise vorticity distribution on the plane of $y/L_r = 2.5375$ whereas invisible on the further upward plane at $y/L_r = 5.075$, where the flow is out of the effect of the primary vortex (see the discussions of Fig.~\ref{fig:plate_222_streamline}). For both cases, the boundary layer is greatly excited downstream the nominal shock impingement location on the side fins around $x/L_r = 192$.
%\begin{figure}[!h]
 %   \centering
%    \includegraphics[width=0.8\textwidth]{eps/Plate_YYY_111_222_vorticity.pdf}
 %   \caption{\RevC{Distributions of the instantaneous streamwise vorticity on the wall-parallel x-z planes at (a) $y/L_r = 2.5375$ and (b) $5.075$.}}
 %   \label{fig:Plate_YYY_111_222_vorticity}
%\end{figure}

\begin{figure}[!h]
    \centering
    \includegraphics[width=0.6\textwidth]{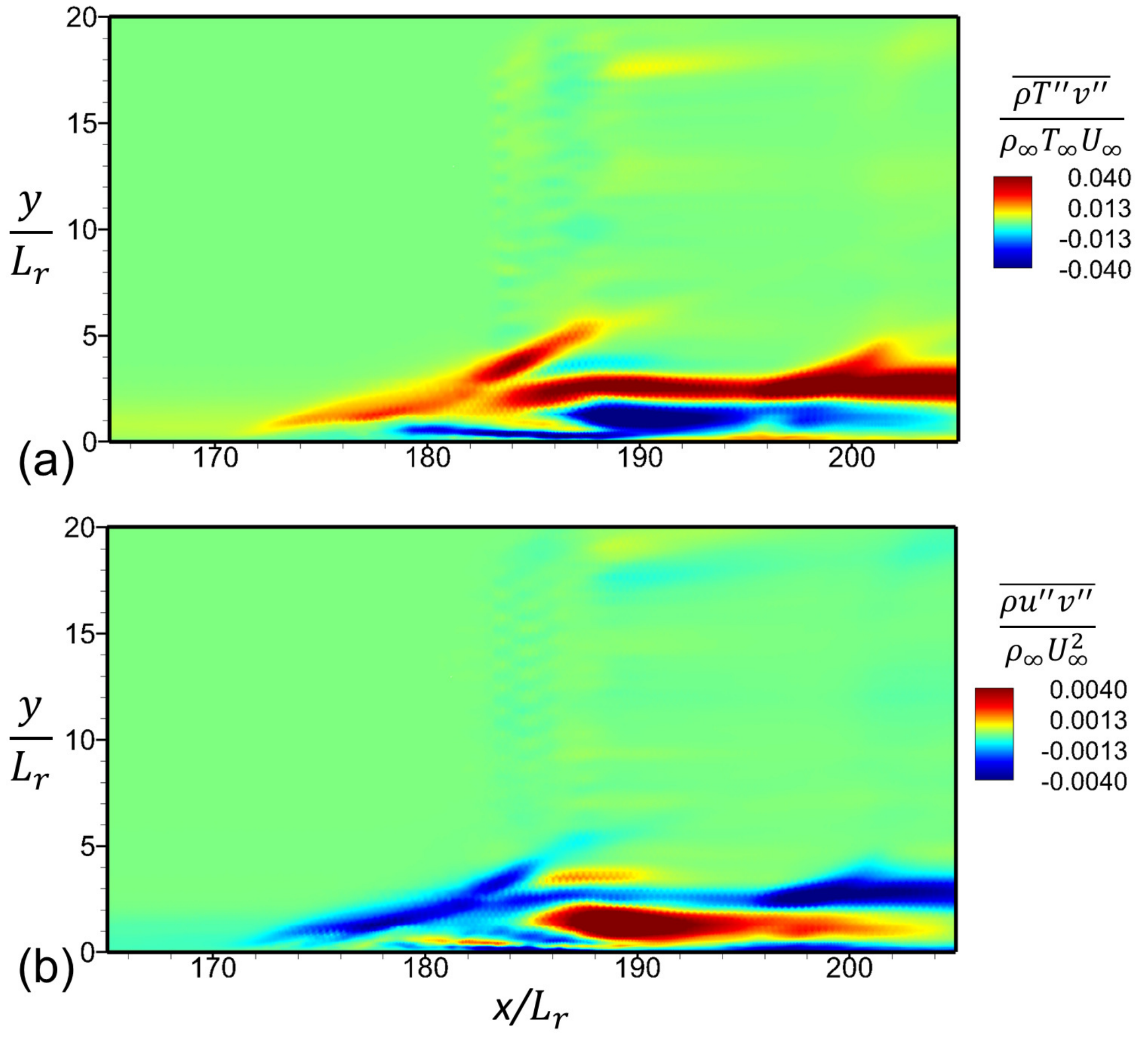}
    \caption{Distributions of (a) the time-averaged wall-normal turbulent heat flux and (b) the time-averaged Reynolds stress on the central wall-normal x-y plane of $z/L_r = 0$.}
    \label{fig:center_plane_reynolds_heat_stress}
\end{figure}

The distributions of the time-averaged wall-normal turbulent heat flux and Reynolds shear stress in the central wall-normal plane are plotted in Fig.~\ref{fig:center_plane_reynolds_heat_stress}. The magnitudes of both quantities grow rapidly at the onset of flow separation at $x/L_r = 175$, and are further amplified by the shock intersection around $x/L_r = 185$. The spatial structures of the average heat flux and Reynolds shear stress are very similar owing to the strong correlation of temperature and streamwise velocity fluctuations (and manifested in the Reynolds Analogy). Both correlations display layered structures with sign reversals. 
%In addition, the primary vortex (see Fig.~\ref{fig:plate_222_streamline}) and the centerline counter-rotating vortex also flip the sign of the wall-normal turbulent heat flux and the Reynolds stress at a certain wall-normal distance, resulting in a layered distribution. 
The impact of this layered distribution on the time-averaged temperature distribution will be discussed in the next section.
\subsubsection{Data analyses in y-z planes}

\begin{figure}[!h]
    \centering
    \includegraphics[width=1\textwidth]{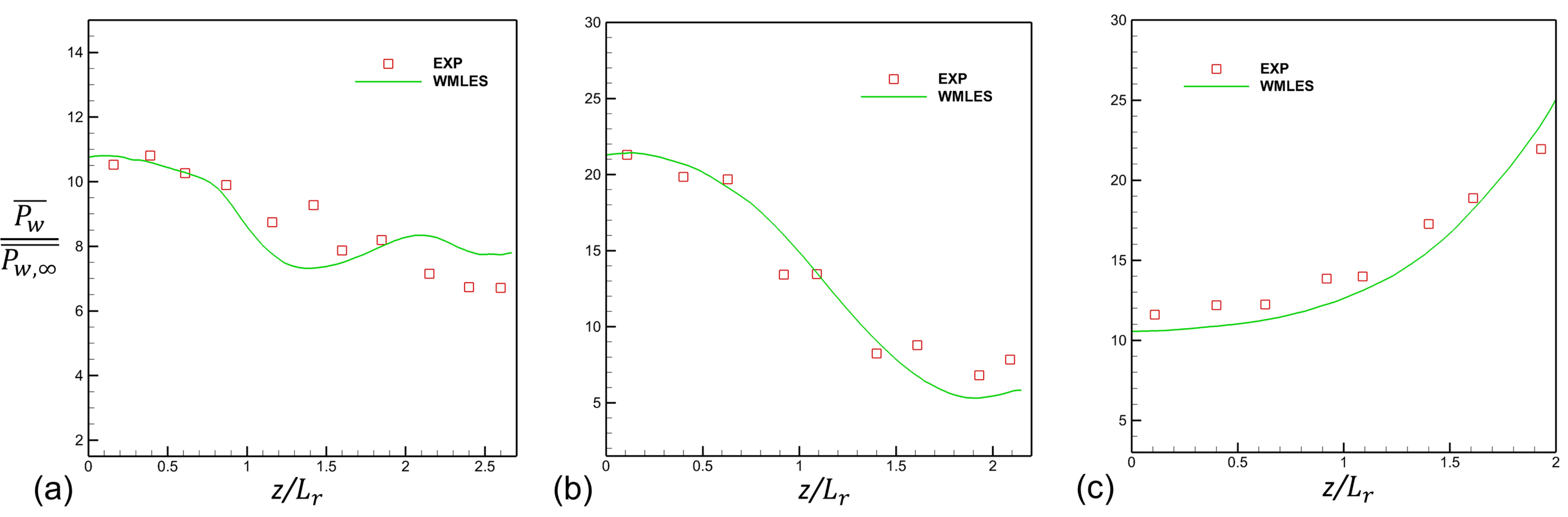}
    \caption{Spanwise distributions of the time-averaged wall pressure $\overline{P_w}/\overline{P_{w,\infty}}$ at the streamwise stations: (a) $x/L_r=183.2$, (b) $187.5$ and (c) $192$. The experimental data (denoted as EXP) are adapted from the Table 4 of \cite{kussoy1992intersecting}.}
    \label{fig:pressure_123_spanwise_compare}
\end{figure}

\begin{figure}[!h]
    \centering
    \includegraphics[width=0.8\textwidth]{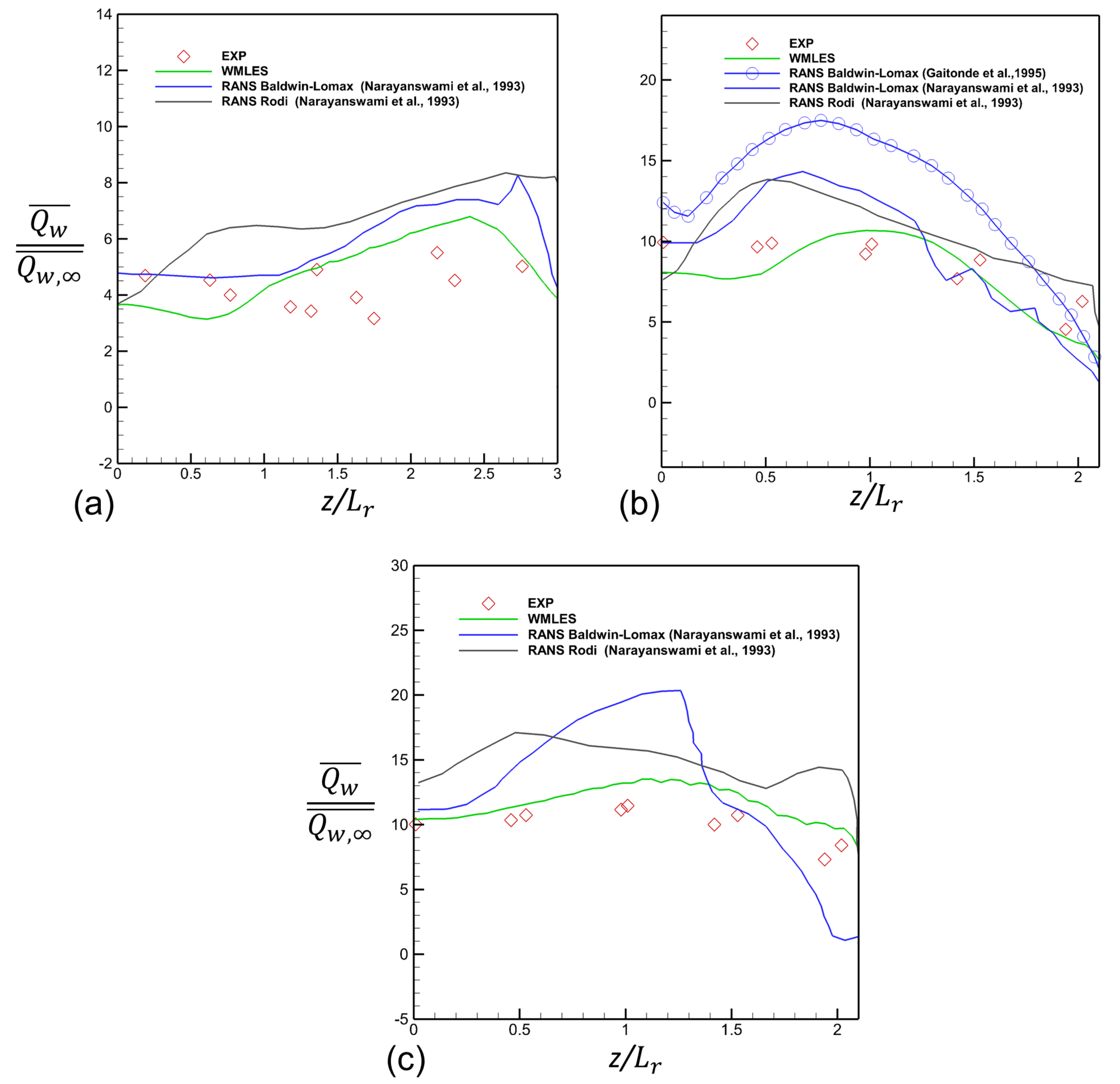}
    \caption{Spanwise distributions of the time-averaged wall heat flux $\overline{Q_w}/\overline{Q_{w,\infty}}$ at the streamwise stations: (a) $x/L_r=181.5$, (b) $185.8$ and (c) $190.3$. The experimental data (denoted as EXP) are from the Table 4 of \cite{kussoy1992intersecting}. In panel (b), the RANS result is computed from the Baldwin-Lomax model with grid E - IB, which is elaborated in Fig.~6 of \cite{gaitonde1995structure}. For all three panels, the RANS predictions in \cite{narayanswami1993numerical} with the zero-equation Baldwin-Lomax model and the two-equation $k-\epsilon$ (Rodi) model are also shown for comparisons. The locations of the spanwise measurements of the surface heat flux are not coincident with those of the surface pressure in Fig.~\ref{fig:pressure_123_spanwise_compare}.}
    \label{fig:heat_123_spanwise_compare}
\end{figure}

The spanwise distributions of the time-averaged surface pressure and surface heat flux at different streamwise stations are given in Fig.~\ref{fig:pressure_123_spanwise_compare} and Fig.~\ref{fig:heat_123_spanwise_compare}, respectively. Considering the reported $10\%$ uncertainties in the experimental data, the spanwise profiles of both quantities are well captured by the present WMLES for all the considered streamwise stations. The time-averaged pressure profile, deviates noticeably  from the experimental measurements in the region $1.2 \le z/L_r \le 2.7$ at station $x/L_r=183.2$ before the shock intersection. The agreement is good at the two further downstream stations. The heat flux distributions are in good agreement with the experimental data and superior to those of RANS predictions. In \cite{narayanswami1993numerical}, it is reported that the peak heat transfer 
in both RANS computations is overestimated by $50\%$ to $75\%$.
%On the other hand, the agreement of the heat flux distribution between the experimental data and the WMLES results is excellent for all three stations. The present results are also better than that from the RANS prediction with the zero-equation Baldwin-Lomax turbulence model \cite{gaitonde1995structure} and that from the two-equation $k-\epsilon$ model \cite{narayanswami1993numerical}. As seen in Fig.~6 of \cite{gaitonde1995structure} and the present Fig.~\ref{fig:heat_123_spanwise_compare}(b), the spanwise heat flux distribution at $x/L_r=185.8$ is substantially over-predicted by the Baldwin-Lomax turbulence model. In Fig.~10 of \cite{narayanswami1993numerical}, it is concluded that both the zero-equation Baldwin-Lomax model and the two-equation $k-\epsilon$ model perform poorly in terms of predicting the transverse profiles of flat plate surface heat transfer for all three streamwise stations.
%

\begin{figure}[!h]
    \centering
    \includegraphics[width=0.45\textwidth]{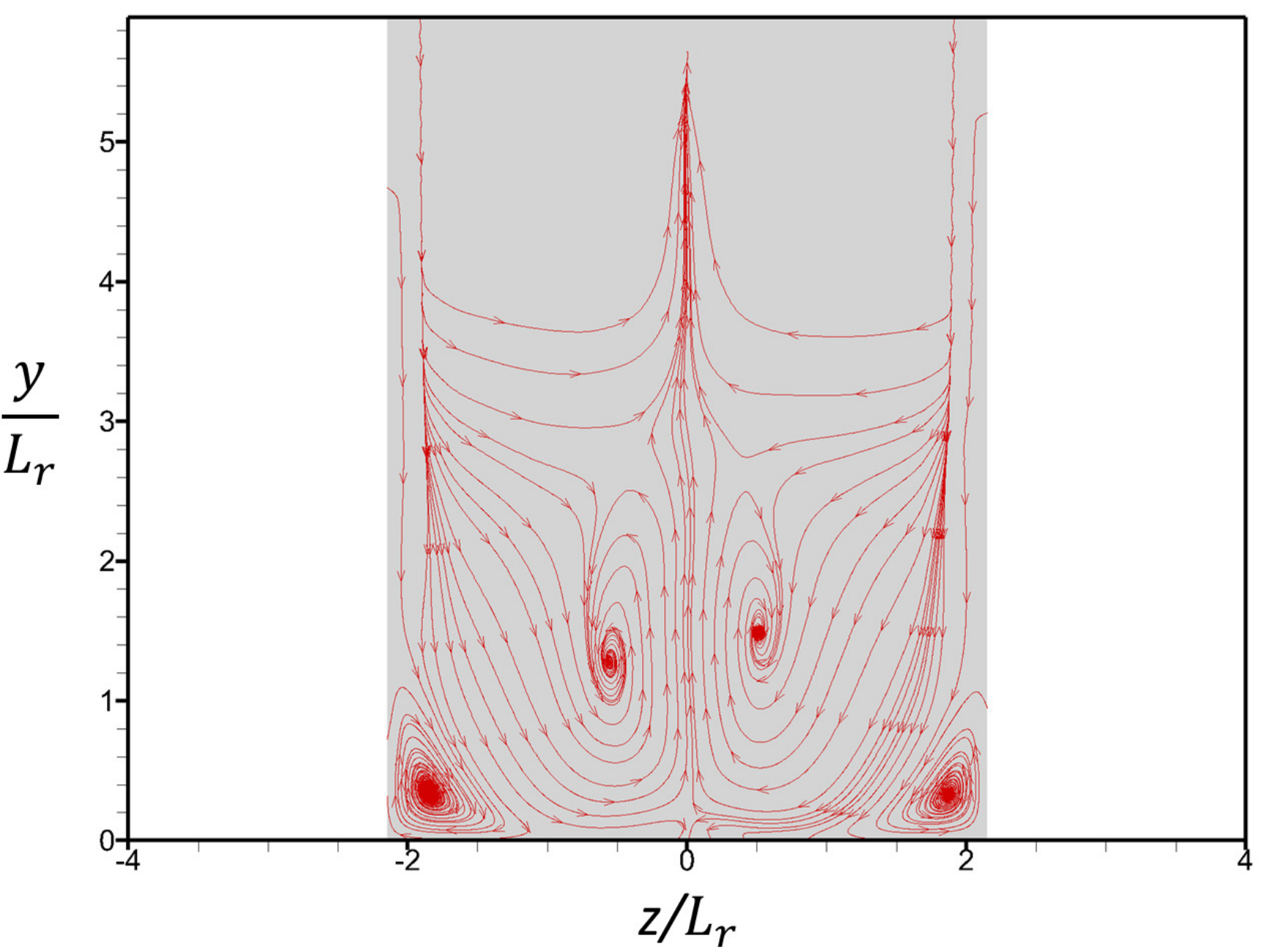}
    \caption{The sectional streamlines on a transverse y-z plane at the streamwise station $x/L_r=187.5$.}
    \label{fig:plate_222_streamline}
\end{figure}

Similar to previous investigations of the crossing shock interaction \cite{narayanswami1993investigation}\cite{gaitonde1995structure}\cite{narayanswami1993numerical}, due to the principal and secondary flow separation analyzed in Fig.~\ref{fig:skin_flat_plate}, the salient feature of the streamline structure is a low total pressure region, accompanied by the primary vortex pair close to the center plane, as shown in Fig.~\ref{fig:plate_222_streamline}.

%%%For instance, on the right side, the counter-clockwise rotating structure around the junction is associated with the corner vortex and the central clockwise rotating vortical structure denotes the primary vortex, which forms the low total pressure region \cite{narayanswami1993numerical}. The third counter-clockwise rotating centerline vortex by the secondary separation is close to the plate and only partially visible due to the coarse LES mesh.

%

\begin{figure}[!h]
    \centering
    \includegraphics[width=0.99\textwidth]{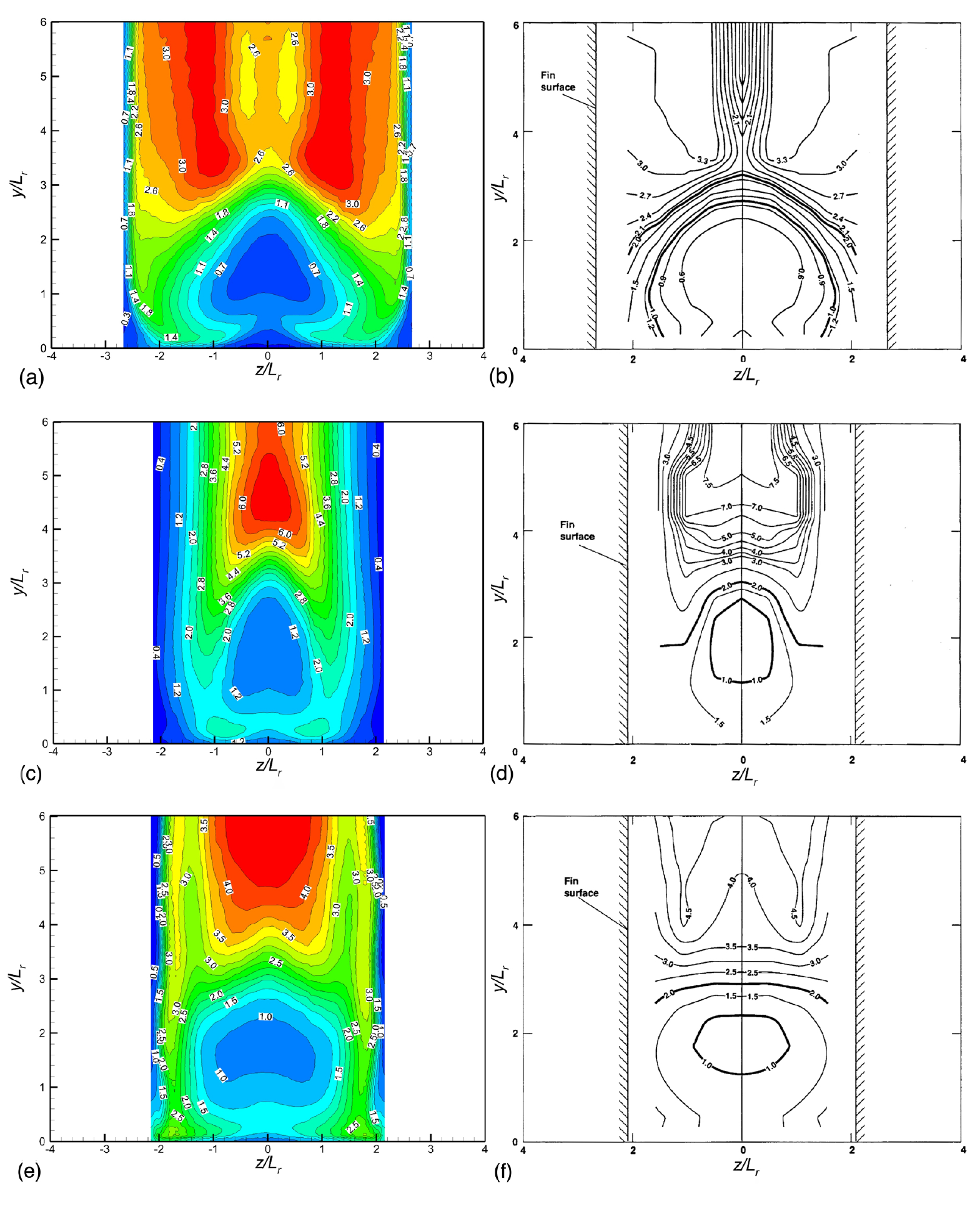}
    \caption{Distribution of the total pressure $\overline{P_\circ}/\overline{P_{\circ,\infty}}$ on a transverse y-z plane at $x/L_r=183.2$ (a,b), $187.5$ (c,d), and $192$ (e,f). Right panels denote the experimental results adapted from Fig.~11 of \cite{kussoy1992intersecting}.}
    \label{fig:Plane_112233_pressure}
\end{figure}

Fig.~\ref{fig:Plane_112233_pressure} shows the total pressure contours on the transverse y-z planes at the pre-shock station, $x/L_r=183.2$, the peak pressure station, $x/L_r=187.5$, and the further downstream station, $x/L_r=192$, respectively. The low total pressure regions are associated with the primary vortex comprising two helical counter-rotating vortices as shown in Fig.~\ref{fig:plate_222_streamline}. 
\RevA{At all the considered streamwise stations, the predicted flow structures are qualitatively similar with those from the experiments. The agreement with the experimental data improves in the post-shock intersection regions, where the influences of inflow conditions are less perceptible. For example, as can be seen in Fig.~\ref{fig:Plane_112233_pressure}(a,b), the total pressure contours from WMLES at $y/L_r \le 3$ have a triangular shape in contrast to the round shape from the experiment at the pre-shock station. This discrepancy (and the effect of inflow conditions) in the pre-shock region is reduced with grid refinement (see Appendix).}
%It is also observed that the low total pressure region tends to gradually move upward from the streamwise station $x/L_r=183.2$ to the downstream station $x/L_r=192$. 
%As shown in Fig.~5 of \cite{narayanswami1993numerical}, the present WMLES predictions of the transverse total pressure profiles are also in better agreement with the experiment than those from both the zero-equation Baldwin-Lomax model and the two-equation $k-\epsilon$ model in the context of RANS approach.
%
%

%
%\begin{figure}[!h]
%    \centering
%    \includegraphics[width=0.85\textwidth]{eps/Plane_222_pressure.pdf}
%    \caption{Distribution of the total pressure $\overline{P_\circ}/\overline{P_{\circ, %\infty}}$ on a transverse y-z plane at $x/L_r=187.5$. Panel (b) denotes the experimental result and is adapted from the Fig.~11(b) of \cite{kussoy1992intersecting}.}
%    \label{fig:Plane_222_pressure}
%\end{figure}
%

%
%\begin{figure}[!h]
%%    \centering
%    \includegraphics[width=0.85\textwidth]{eps/Plane_333_pressure.pdf}
%    \caption{Distribution of the total pressure $\overline{P_\circ}/\overline{P_{\circ, \infty}}$ on a transverse y-z plane at $x/L_r=192$. Panel (b) denotes the experimental result and is adapted from the Fig.~11(c) of \cite{kussoy1992intersecting}.}
%    \label{fig:Plane_333_pressure}
%\end{figure}
%

\begin{figure}[!h]
    \centering
    \includegraphics[width=1\textwidth]{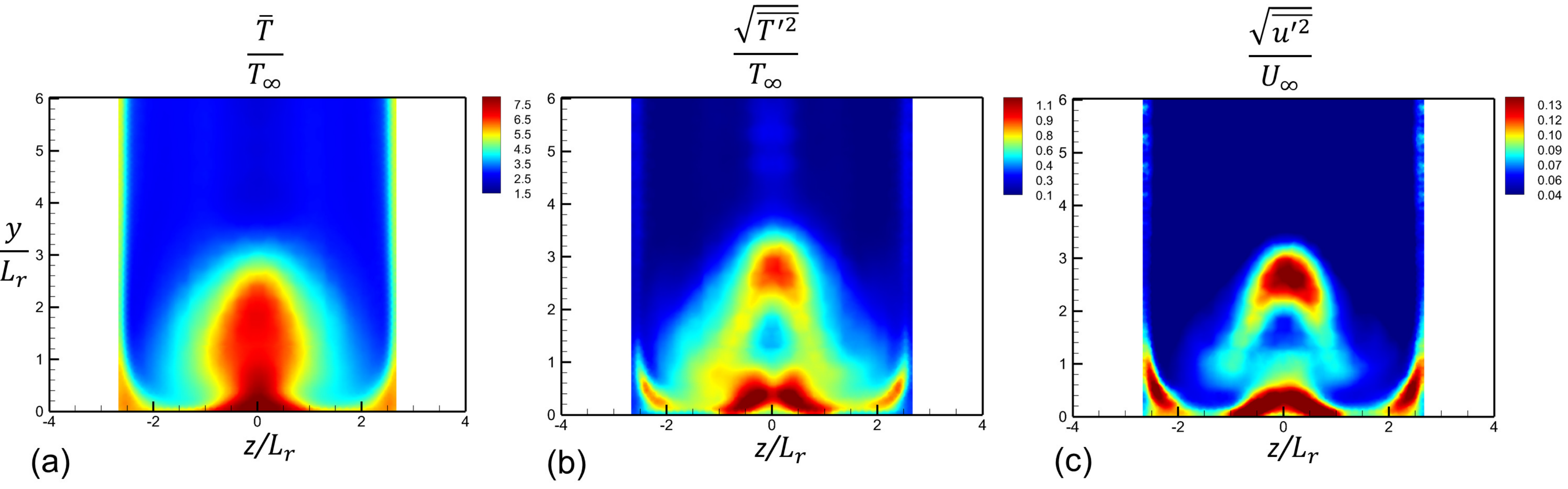}
    \includegraphics[width=1\textwidth]{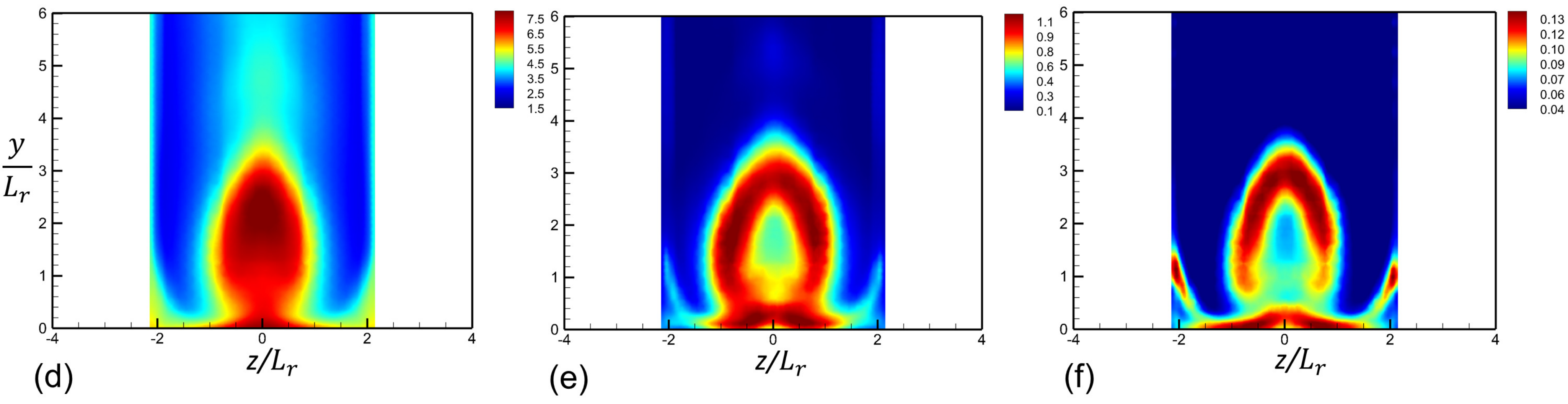}
    \includegraphics[width=1\textwidth]{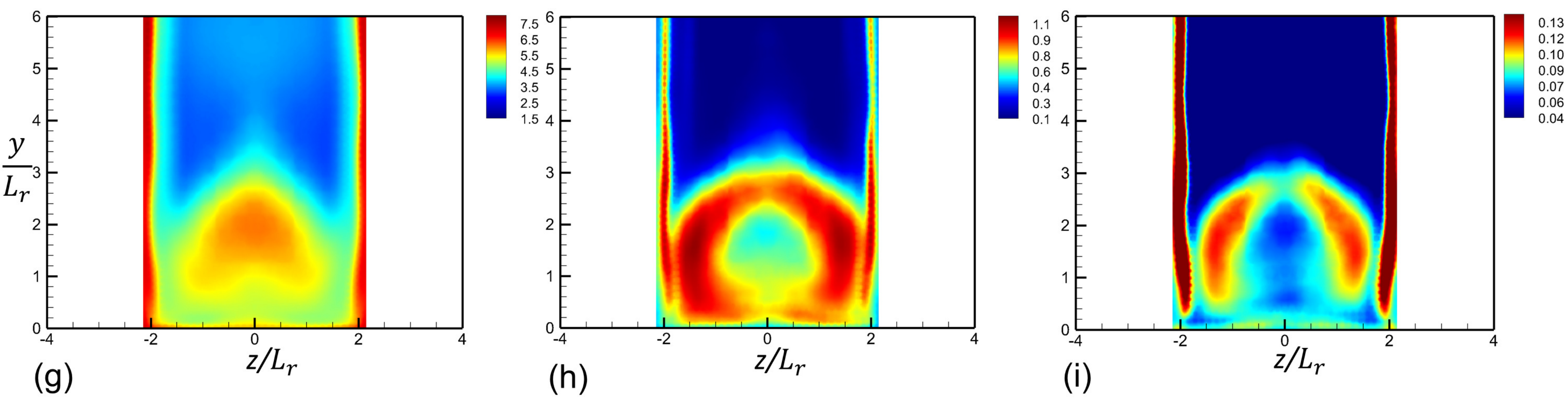}
    \caption{Distributions of the time-averaged temperature (a,d,g), the r.m.s temperature fluctuations (b,e,h), and the r.m.s streamwise velocity fluctuations (c,f,i) on the transverse y-z planes at the streamwise stations with $x/L_r=183.2$ (a,b,c), $187.5$ (d,e,f), and $192$ (g,h,i).}
    \label{fig:123_TU}
\end{figure}

Fig.~\ref{fig:123_TU} shows the distributions of the time-averaged temperature, r.m.s temperature fluctuations, and r.m.s streamwise velocity fluctuations at $x/L_r=183.2$, $187.5$, and $192$. All the quantities shown peak along the central region between the fins. The peak mean temperature is reached away from the flat plate and in the shock intersection region where the velocity and temperature fluctuations are suppressed. As the flow develops further downstream, regions of higher mean temperature and intense velocity and temperature fluctuations move closer to the fins.

%Before the shock intersection, at station $x/L_r=183.2$, the boundary layer is dramatically lifted up and filled with hot fluids due to the primary vortex, and the centerline vortex induced by the shock/boundary-layer interaction on the flat plate. Due to the strong counter-rotating vortices near the centerline, the peak temperature appears close to the flat plate, where the peak temperature and velocity fluctuations also occur. Immediately downstream of the shock intersection, at station $x/L_r=187.5$, the peak temperature region further moves upward while the overall flame-like structure of the temperature distribution remains a similar shape. Conversely, the peak temperature and streamwise velocity fluctuations show up on the outer edge of the flame-like structure, and the effects of the centerline secondary separation become weaker. The prominent temperature and streamwise velocity fluctuations also correlate with the unsteadiness of the off-plate primary vortex after the shock intersection. Further downstream at $x/L_r=192$, the effects of the secondary separation almost disappear and the flame-like structure becomes flatter with a significant amount of hot fluids accumulated within the thin layers attached to the side fins. This redistribution of hot fluids is probably due to the presence of the separation induced by the shock impingement on the fin surfaces, see also Fig.~\ref{fig:skin_friction_fin}.
%

\begin{figure}[!h]
    \centering
    \includegraphics[width=0.75\textwidth]{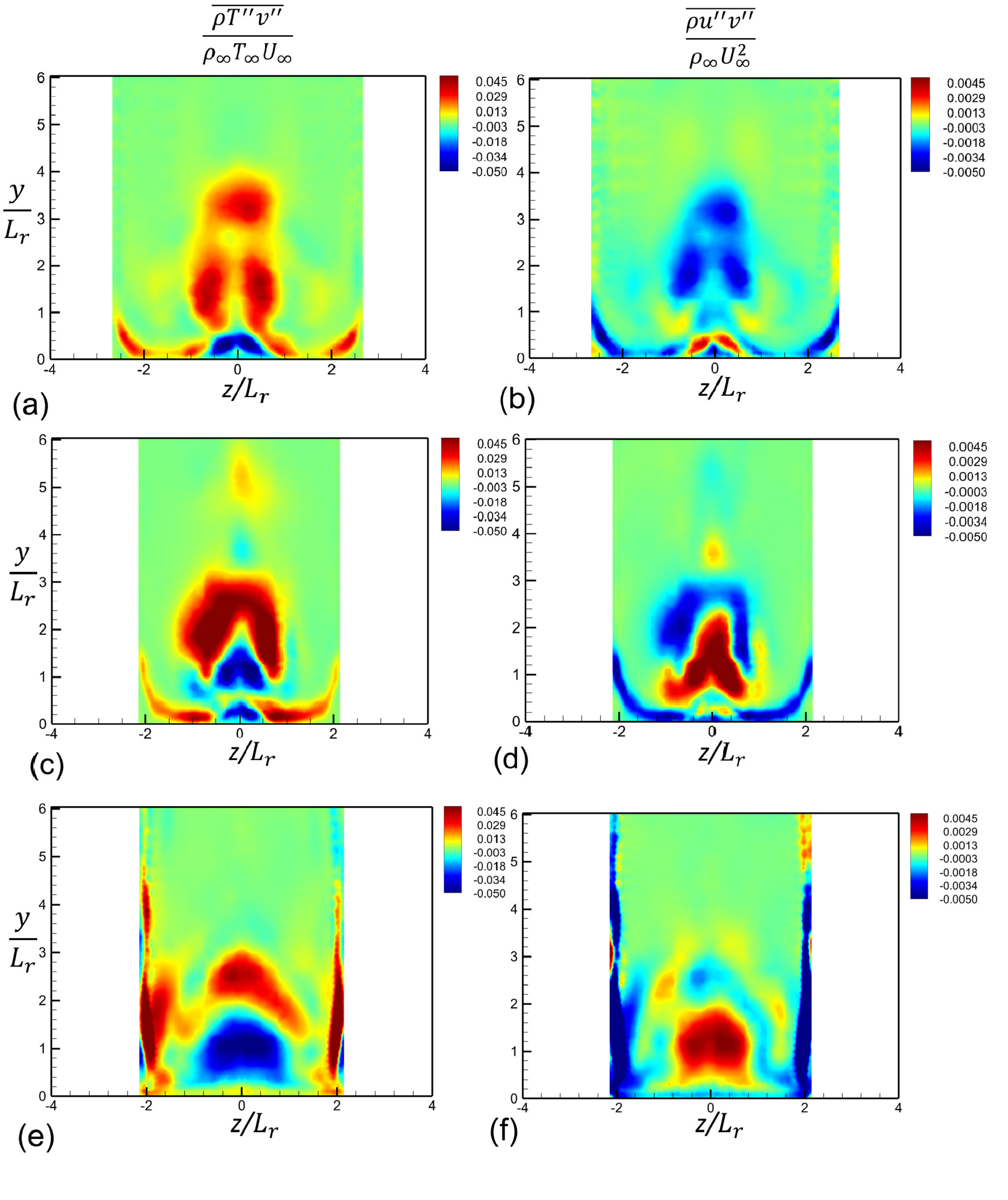}
    \caption{\RevC{Distributions of the time-averaged wall-normal turbulent heat flux (a,c,e), and the time-averaged Reynolds stress (b,d,f) on the transverse y-z planes at the streamwise station $x/L_r=183.2$ (a,b), $187.5$ (c,d), and $192$ (e,f).}}
    \label{fig:123_reynolds_heat_stress}
\end{figure}

Fig.~\ref{fig:123_reynolds_heat_stress} shows the time-averaged turbulent heat flux and Reynolds stress on the transverse y-z planes at three streamwise stations. Once again, a strong correlation between the Reynolds shear stress and heat flux is apparent. Knowledge of the spatial structure of these correlations is valuable in RANS turbulence modeling, where both correlations are phenomenologically modeled in the governing equations for the mean velocity and temperature. The sign reversals of heat flux (and Reynolds stress) in the transverse planes displayed 
are a consequence of the streamwise vortices and flow reversals owing to the intersecting shocks. \RevC{The predicted flow fields are marginally asymmetric mainly due to the staggered nature of the deployed unstructured Voronoi mesh and the coarse resolution. Particularly for the downstream unsteady regions, the flow symmetry is more sensitive to the mesh topology. Note that the mean velocity and temperature profiles, and normal components of turbulent intensities are nearly symmetric, but the cross correlations may require much longer time averaging and are apparently more susceptible to asymmetries in the unstructured mesh.  } 
%Note that the flow symmetry is not improved with longer time-averaging.}

%%Note that, the correlations are not symmetric with respect to the center plane between the fins. Generally, cross correlations require larger statistical samples (longer integration times) for convergence than normal intensities.  

%Taking the distribution of the wall-normal turbulent heat flux at $x/L_r=187.5$ for instance, along the centerline of $z/L_r = 0$, $\overline{\rho T''v''}$ remains negative pushing the hot fluids downward until approximately $y/L_r \approx 2.2$. Conversely, $\overline{\rho T''v''}$ becomes positive lifting the hot fluids up for the region with $y/L_r > 2.2$. Consequently, the temperature peak is produced around $y/L_r \approx 2.2$, which is consistent with the time-averaged temperature distribution as in Fig.~\ref{fig:123_TU}(d). The similar analyses are applicable for interpreting the time-averaged temperature distribution at other streamwise stations. 
%
\subsection{WMLES with van Driest scaling based damping function}\label{sect:vD_scaling}

\begin{figure}[!h]
    \centering
    \includegraphics[width=0.9\textwidth]{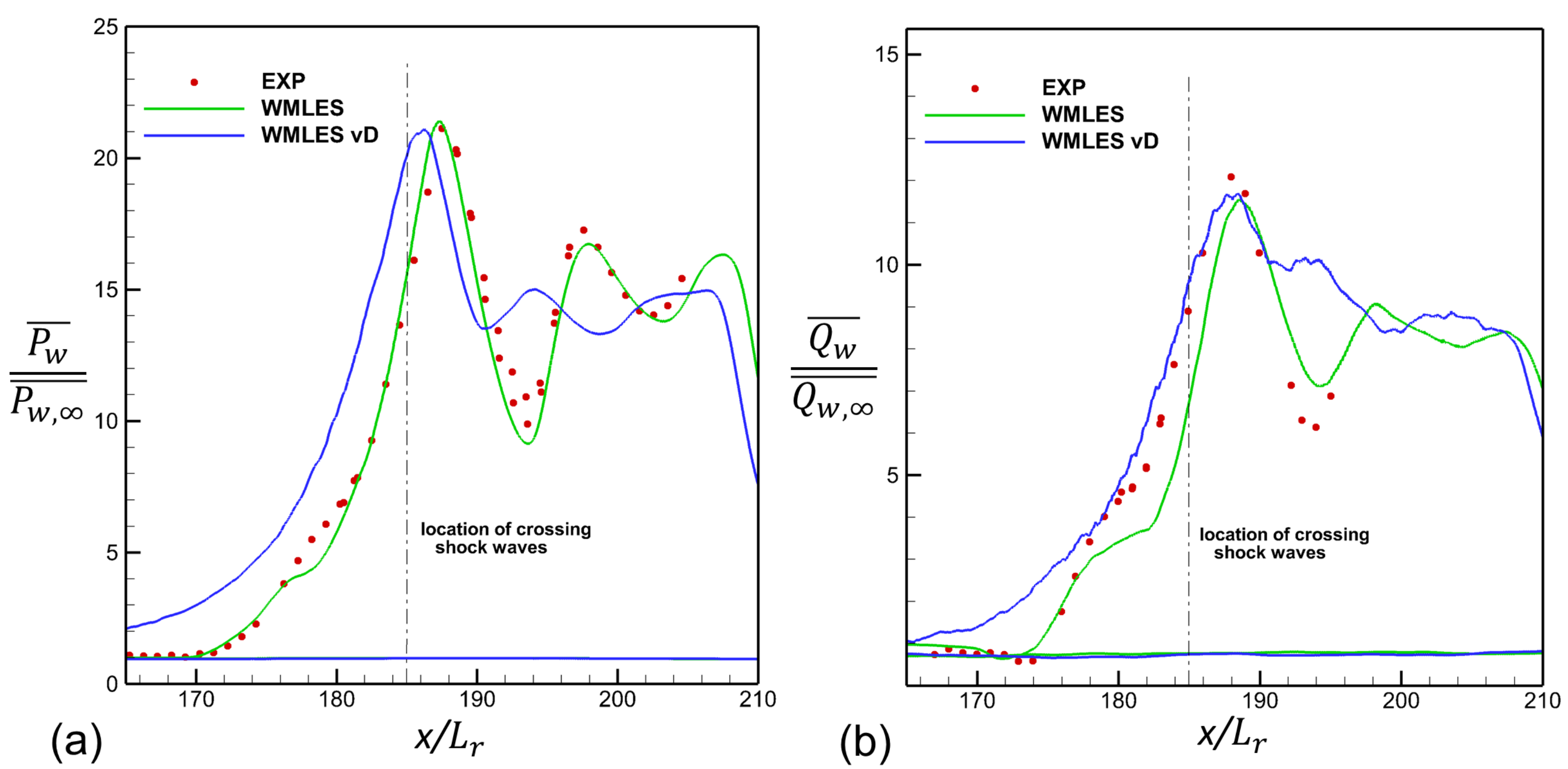}
    \caption{Streamwise distributions of the time-averaged (a) surface pressure and (b) surface heat flux on the flat plate at  $z/L_r = 0$. The green lines, the blue lines and the red symbols denote the data from the WMLES with the semi-local scaling, the WMLES with van Driest scaling, and the experiment, respectively.}
    \label{fig:VD_DATA_compare_streamwise}
\end{figure}

\begin{figure}[!h]
    \centering
    \includegraphics[width=0.9\textwidth]{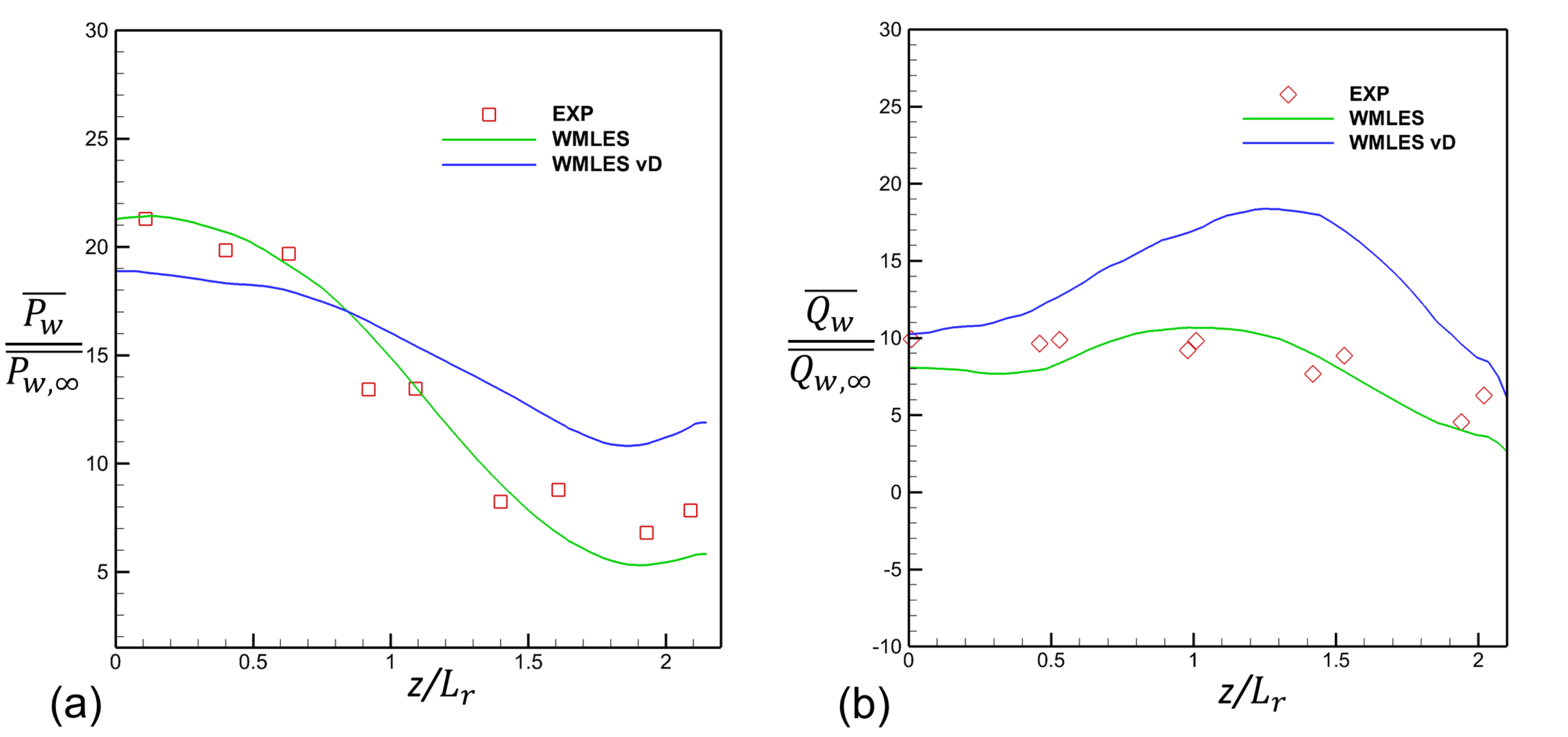}
    \caption{Spanwise distributions of (a) the time-averaged surface pressure at the streamwise station $x/L_r=187.5$ and (b) the time-averaged surface heat flux at the streamwise station $x/L_r=185.8$. The green lines, the blue lines and the red symbols denote the data from the WMLES with the semi-local scaling, the WMLES with the van Driest scaling, and the experiment, respectively.}
    \label{fig:VD_DATA_compare_spanwise}
\end{figure}

In this section, the sensitivity of the results to the coordinate scaling of the wall model eddy viscosity is further evaluated.  Wall modeled LES calculations on the ``fine'' grid using identical freestream boundary conditions described above were additionally performed using the van Driest damping function \cite{kawai2012wall}. Comparison of the time-averaged pressure and heat flux distributions between the WMLES with the van Driest scaling and semi-local scaling is provided in Fig.~\ref{fig:VD_DATA_compare_streamwise} and Fig.~\ref{fig:VD_DATA_compare_spanwise}. For all the concerned quantities, the accuracy of the WMLES deteriorates when the van Driest scaling is deployed in the damping function in the eddy viscosity model. In particular, both the mean pressure and heat flux are notably over-predicted near the entrance of the double fins. Distributions of both quantities are also poorly captured after the shock intersection. Comparisons of the spanwise profiles of average pressure and surface heat flux with the experimental data also show higher accuracy with the semi-local scaling.

\begin{figure}[!h]
    \centering
    \includegraphics[width=0.8\textwidth]{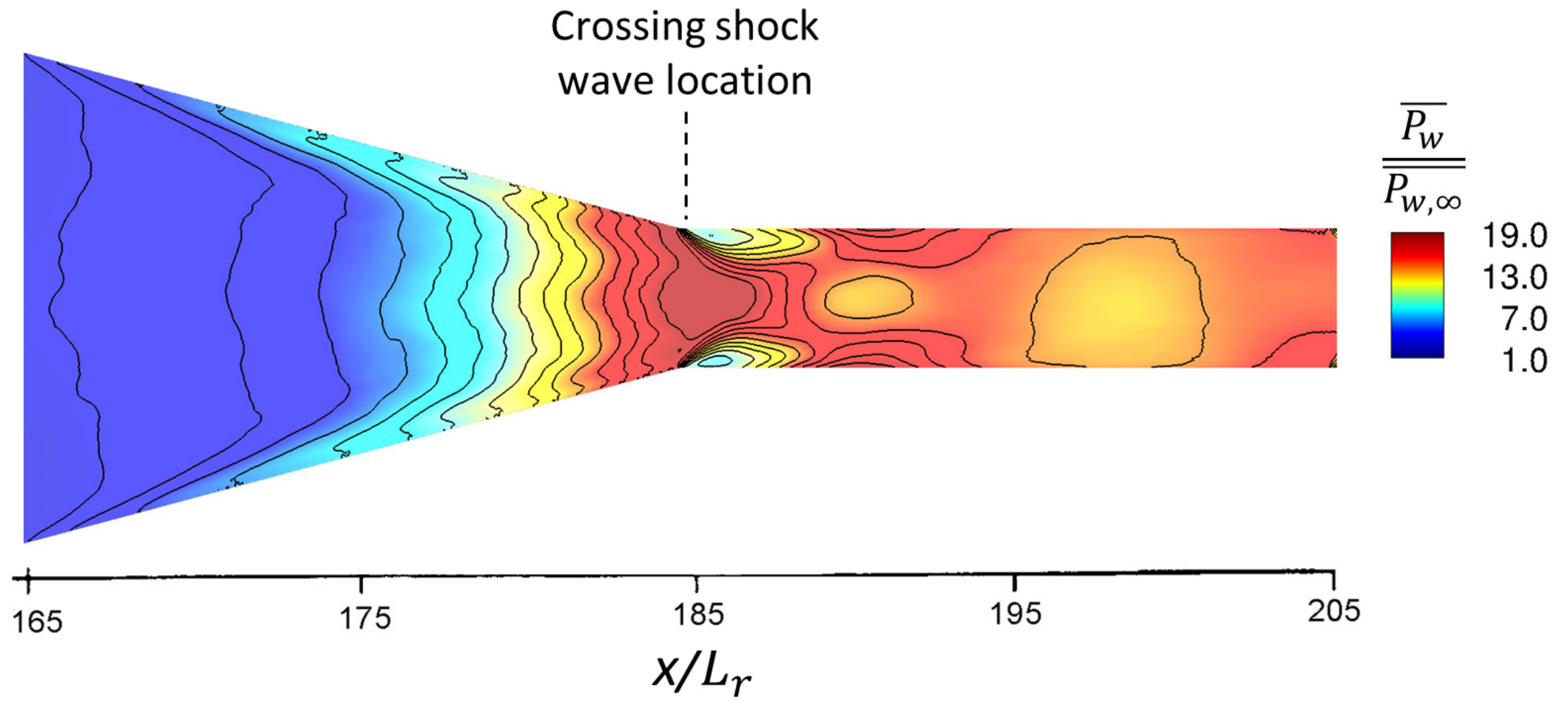}
    \caption{Distribution of the time-averaged pressure on the flat-plate surface at $y/L_r = 0$. The location of the double-shock intersection based on the inviscid theory is also shown in the plot. The result reported in this plot is from WMLES with van Driest scaling based damping function.}
    \label{fig:VD_pressure_plate}
\end{figure}

As shown in Fig.~\ref{fig:VD_pressure_plate}, the structure of wall pressure distribution is significantly different from that with semi-local scaling in Fig.~\ref{fig:pressure_heat_plate}(a). 
%This result is consistent with the observations that semi-local scaling better collapses compressible velocity profiles to their incompressible counterparts compared to the van Driest transformation, particularly in the viscous sublayer with isothermal wall conditions \cite{zhang2018direct}\cite{iyer2019analysis}\cite{Yang2017eddy}.

%

\begin{figure}[!h]
    \centering
    \includegraphics[width=0.8\textwidth]{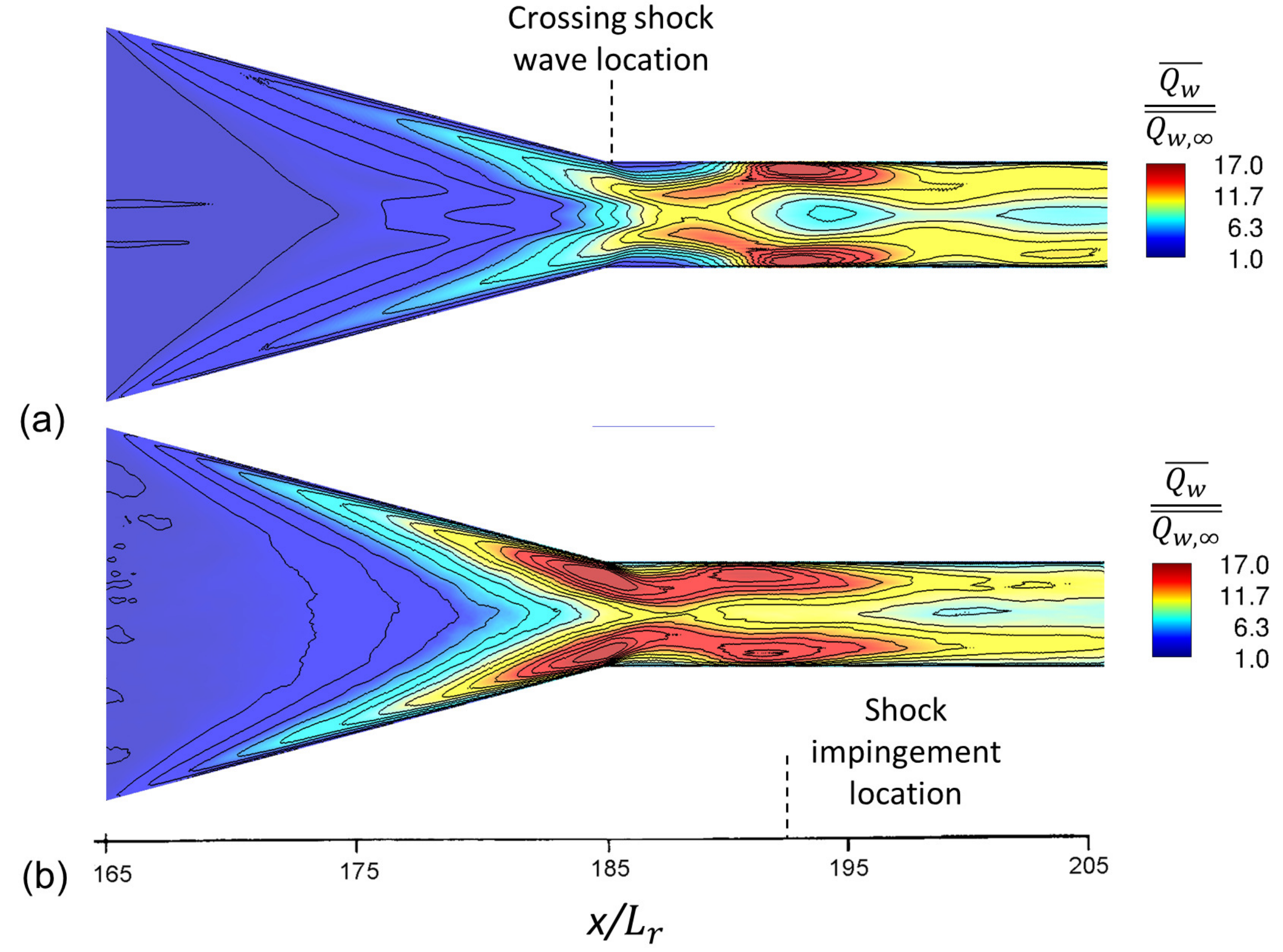}
    \caption{\RevA{Distributions of the time-averaged surface heat flux from WMLES with (a) the semi-local scaling and (b) the van Driest scaling on the flat plate at $y/L_r = 0$. The location of the double-shock intersection based on the inviscid theory is shown in the plots. Also shown is the shock impingement location around $x/L_r = 192$.}}
    \label{fig:heat_plate_SL_VD_compare}
\end{figure}

\RevA{In Fig.~\ref{fig:heat_plate_SL_VD_compare}, the surface heat flux distributions predicted with WMLES with the semi-local scaling and the van Driest scaling are compared. The notable differences present around the centerline secondary separation, the fin corner regions, and the regions right downstream of the shock intersection. The secondary separation is not well captured by the van Driest scaling, which can also be confirmed in Fig.~\ref{fig:VD_DATA_compare_streamwise}(b), where the plateau is missed around $x/L_r = 180$. In the fin corner regions around $x/L_r = 185$, where the expansion dominates, the surface heat flux is significantly overpredicted when compared to the semi-local scaling. Downstream of the shock intersection, the valley in the heat flux is also missed by the van Driest scaling. }

%Recalling that the heat flux prediction from the van Driest scaling is quantitatively erroneous even for canonical zero-pressure-gradient boundary layer flows, the current investigation shows that the prediction discrepancy is more prominent in the regions where strong nonequilibrium presents.
%
\section{Conclusions}

In this study, an experimentally well documented hypersonic inlet flow involving complex three-dimensional intersecting shock-wave/turbulent boundary-layer interaction and flow separation, is investigated using wall modeled large eddy simulation. Despite the presence of complex non-equilibrium phenomena, the results from WMLES with equilibrium wall model (also with the low-dissipation numerical method and the high-quality Voronoi mesh in the charLES solver) agree favorably with experimental data for mechanical loading, surface heat fluxes, and for the prediction of the secondary separation in both the shock intersection and post-shock regimes. The use of the semi-local scaling in the eddy viscosity of the wall model leads to significant improvements in the results compared to the van Driest scaling. The WMLES predictions are shown to be significantly more accurate than those of prior RANS calculations using either Baldwin-Lomax or $k-\epsilon$ models. The coarseness of the WMLES calculations (relative to the boundary layer thickness or size of the separation bubble) suggest that this approach (which consists of a unique combination of accurate numerical methodology and LES) can be affordable for high speed aerodynamics simulations in geometries of engineering interest.

\section*{Acknowledgments}
This work was supported by NASA under grant number NNX15AU93A, and AFOSR under grant number FA9550-16-1-0319.
Supercomputing resources were provided through the INCITE Program of the Department of Energy (DOE). Mori Mani and Matthew Lakebrink from Boeing Research $\&$ Technology are acknowledged for suggesting this case to the authors. The first author appreciates useful discussions with Kevin Griffin at CTR, Stanford University.

\section*{Data availability}
The data that support the findings of this study are available on request from the corresponding author, LF.

\section*{Appendix A. Statistical and grid convergence}\label{sect:convergence_study}

In this section, the statistical and grid convergence of the main quantities of interest are investigated.

\subsection*{A1. Averaging time convergence study}\label{sect:time_study}

\begin{figure}[!h]
    \centering
    \includegraphics[width=1.\textwidth]{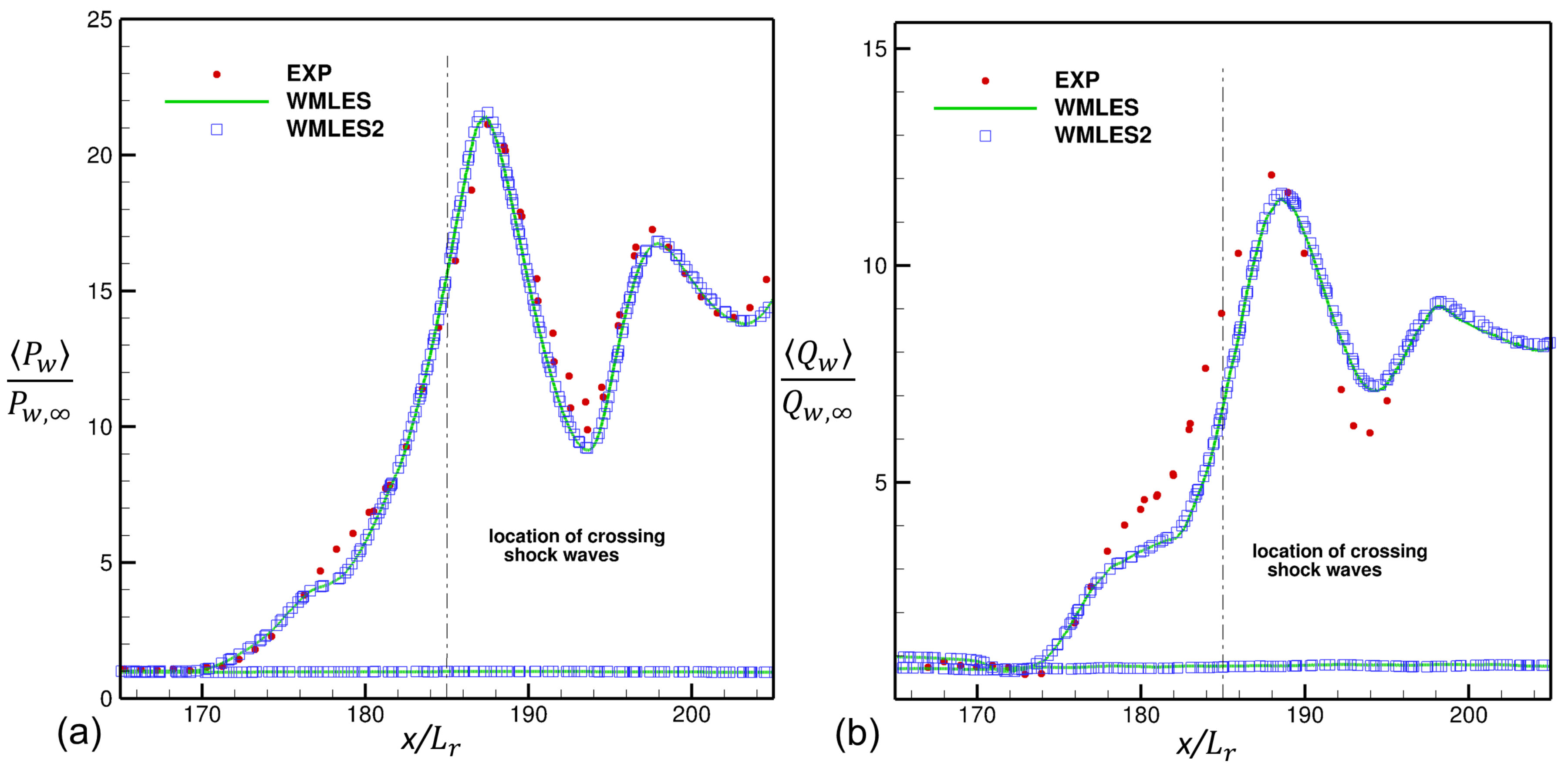}
    \caption{\RevC{Streamwise distributions of the time-averaged (a) surface pressure and (b) surface heat flux on the flat plate at $z/L_r = 0$. WMLES denotes the result reported in Fig.~\ref{fig:double_fins_pressure_heat} and WMLES2 denotes the solution averaged in a time interval, which is 8 flow-through times longer than that of WMLES.}}
    \label{fig:time_convergence}
\end{figure}

\RevB{As shown in Fig.~\ref{fig:time_convergence}, increasing the time averaging interval by 8 flow through times does not affect the pressure and mean surface heat flux statistics, and hence these key quantities of practical interest are considered statistically converged.}

%As discussed earlier, some higher order statistical correlations such as turbulent heat flux and Reynolds shear stress require longer integration times for statistical convergence.
%
\subsection*{A2. Resolution sensitivity study}\label{sect:resolution_study}

\begin{figure}[!h]
    \centering
    \includegraphics[width=1.\textwidth]{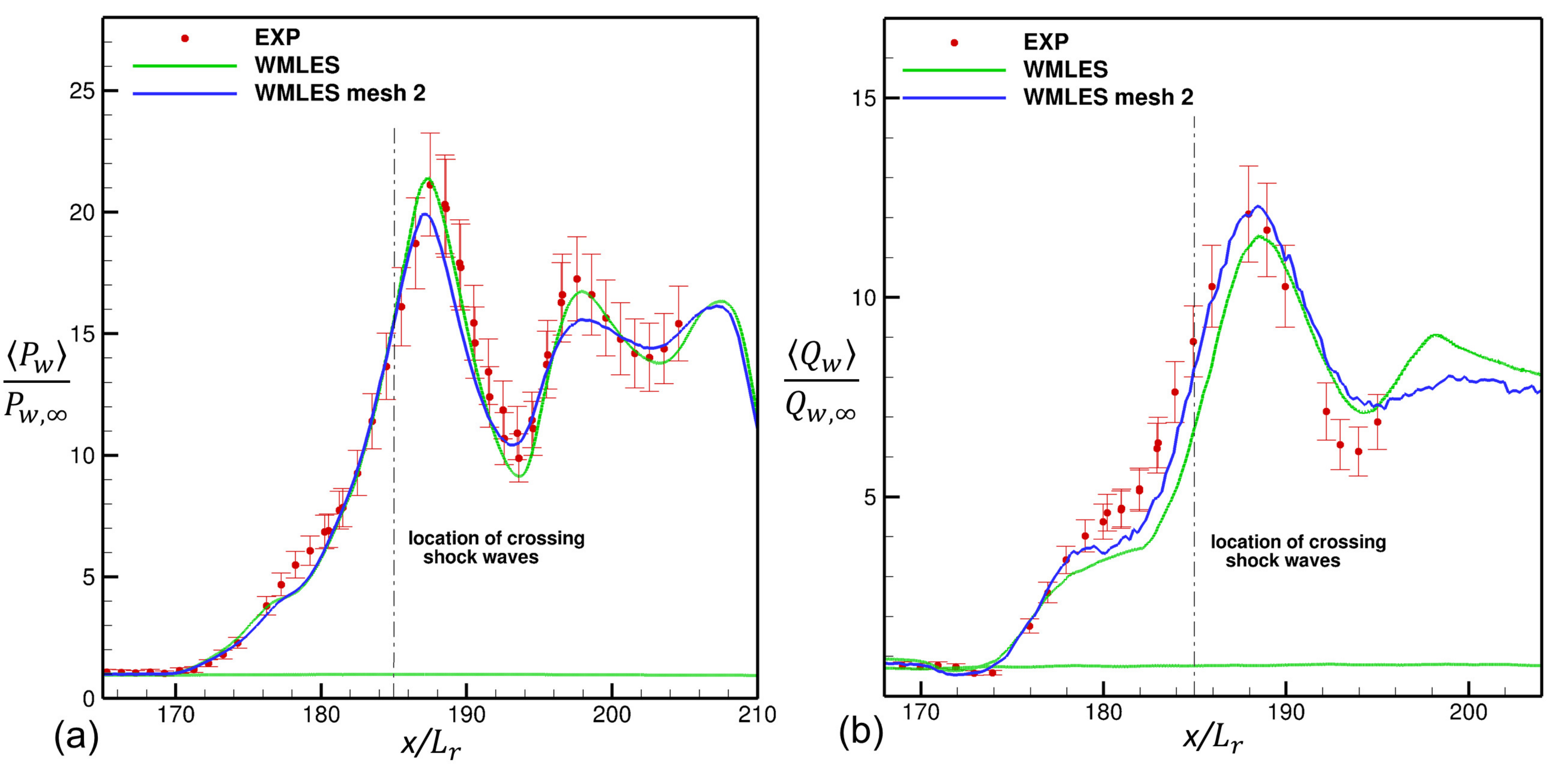}
    \caption{\RevB{Streamwise distributions of the time-averaged (a) surface pressure and (b) surface heat flux on the flat plate at $z/L_r = 0$. The results from WMLES with 70M cells and WMLES mesh 2 with 143M cells are reported for comparisons. Also plotted are the uncertainty bars reported from the experiment.}}
    \label{fig:resolution_convergence}
\end{figure}

\begin{figure}[!h]
    \centering
    \includegraphics[width=0.99\textwidth]{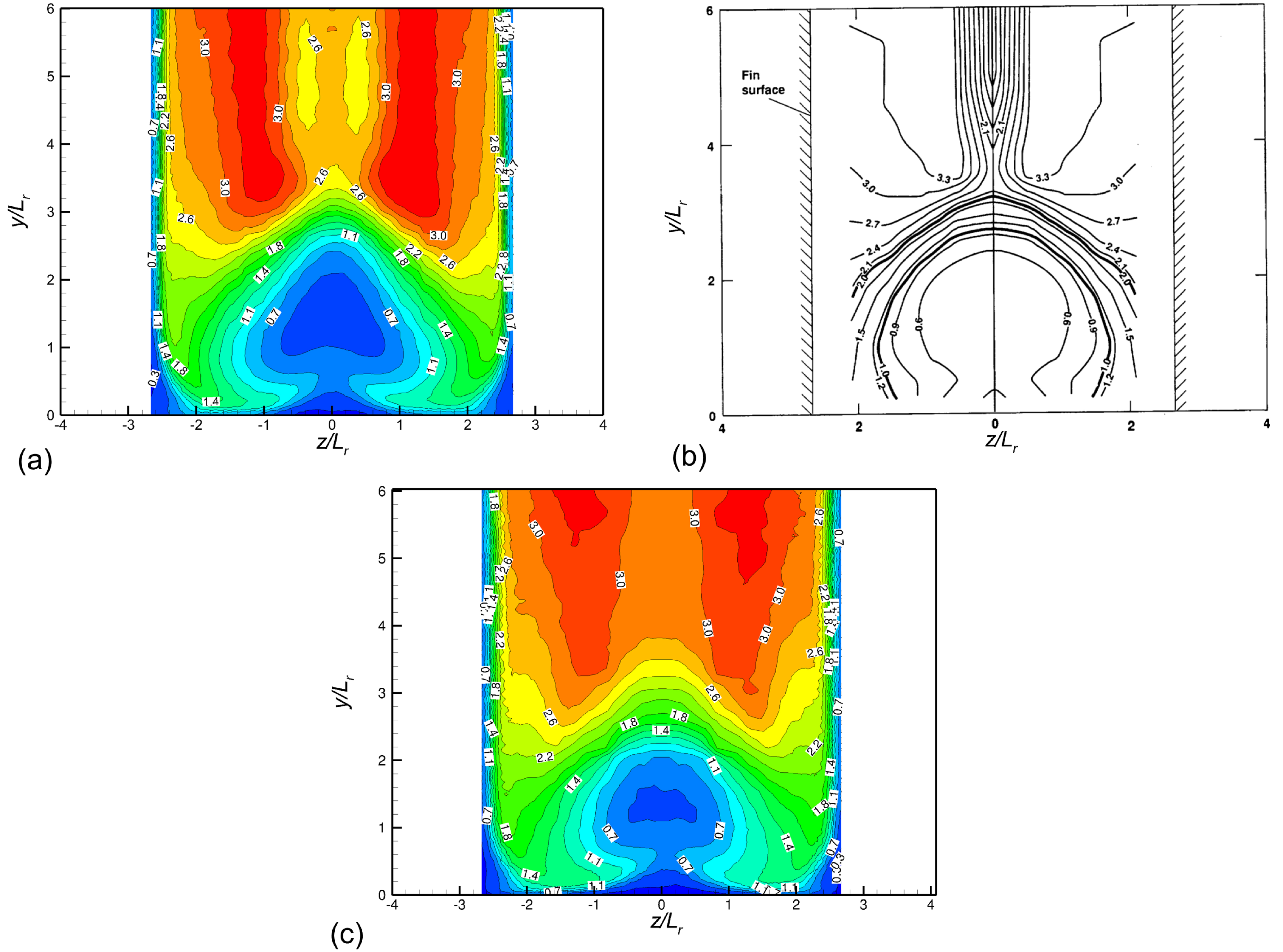}
    \caption{\RevA{Distribution of the total pressure $\overline{P_\circ}/\overline{P_{\circ,\infty}}$ on a transverse y-z plane at $x/L_r=183.2$. Panel (b) denotes the experimental result and is adapted from the Fig.~11(a) of \cite{kussoy1992intersecting}. Panel (a) and (c) denote the WMLES results from the mesh with 70M cells and 143M cells, respectively.}}
    \label{fig:resolution_plane_111_pressure}
\end{figure}

\RevB{A higher-resolution simulation with 143M cells was carried out. This mesh is generated by refining the near-wall region of the mesh with 70M cells (as described in Table~\ref{tab:double_fins_grid}). As shown in Fig.~\ref{fig:resolution_convergence}, the results from both resolutions are generally within the experimental uncertainty bars of measured  wall pressure. The heat flux predictions upstream of $x/L_r=190$ are improved with the finer mesh.} \RevA{In terms of the flow structure, as shown in Fig.~\ref{fig:resolution_plane_111_pressure}, the shape of the predicted separation bubble from the higher resolution agrees with the experimental sketch better.
Further mesh refinements, especially in the vicinity of the separation bubble may improve the predictions. However, given the intrinsic uncertainties in the prescription of inflow conditions, and the higher cost of more refined computations, we did not carry out additional simulations with finer grid resolution. As remarked earlier, the LES results are always going to be grid dependent, but do converge to DNS in the limit of very fine grids. Here, we have demonstrated the level of accuracy that can be expected at affordable cost.}

%

%% References


\bibliographystyle{spmpsci}      % mathematics and physical sciences

\bibliography{explicit}

\begin{thebibliography}{10}
\providecommand{\url}[1]{{#1}}
\providecommand{\urlprefix}{URL }
\expandafter\ifx\csname urlstyle\endcsname\relax
  \providecommand{\doi}[1]{DOI~\discretionary{}{}{}#1}\else
  \providecommand{\doi}{DOI~\discretionary{}{}{}\begingroup
  \urlstyle{rm}\Url}\fi

\bibitem{adams2000direct}
Adams, N.A.: {Direct simulation of the turbulent boundary layer along a
  compression ramp at $M$=3 and $Re_\theta$= 1685}.
\newblock Journal of Fluid Mechanics \textbf{420}, 47--83 (2000)

\bibitem{aurenhammer1991voronoi}
Aurenhammer, F.: {Voronoi diagrams--a survey of a fundamental geometric data
  structure}.
\newblock ACM Computing Surveys (CSUR) \textbf{23}(3), 345--405 (1991)

\bibitem{baldwin1978thin}
Baldwin, B., Lomax, H.: {Thin-layer approximation and algebraic model for
  separated turbulentflows}.
\newblock In: 16th aerospace sciences meeting, p. 257 (1978)

\bibitem{bermejo2014confinement}
Bermejo-Moreno, I., Campo, L., Larsson, J., Bodart, J., Helmer, D., Eaton,
  J.K.: {Confinement effects in shock wave/turbulent boundary layer
  interactions through wall-modelled large-eddy simulations}.
\newblock Journal of Fluid Mechanics \textbf{758}, 5--62 (2014)

\bibitem{bose2018wall}
Bose, S.T., Park, G.I.: {Wall-modeled large-eddy simulation for complex
  turbulent flows}.
\newblock Annual Review of Fluid Mechanics \textbf{50}, 535--561 (2018)

\bibitem{bres2018large}
Bres, G.A., Bose, S.T., Emory, M., Ham, F.E., Schmidt, O.T., Rigas, G.,
  Colonius, T.: Large-eddy simulations of co-annular turbulent jet using a
  {V}oronoi-based mesh generation framework.
\newblock In: 2018 AIAA/CEAS Aeroacoustics Conference, p. 3302 (2018)

\bibitem{bres2019modelling}
Br{\`e}s, G.A., Lele, S.K.: {Modelling of jet noise: a perspective from
  large-eddy simulations}.
\newblock Philosophical Transactions of the Royal Society A \textbf{377}(2159),
  20190081 (2019)

\bibitem{choi2012grid}
Choi, H., Moin, P.: {Grid-point requirements for large eddy simulation:
  Chapman's estimates revisited}.
\newblock Physics of fluids \textbf{24}(1), 011702 (2012)

\bibitem{currao2019hypersonic}
Currao, G.M., Choudhury, R., Gai, S.L., Neely, A.J., Buttsworth, D.R.:
  {Hypersonic Transitional Shock-Wave--Boundary-Layer Interaction on a Flat
  Plate}.
\newblock AIAA Journal pp. 1--16 (2019)

\bibitem{duan2011directMach}
Duan, L., Beekman, I., Martin, M.P.: Direct numerical simulation of hypersonic
  turbulent boundary layers. {P}art 3. {E}ffect of {M}ach number.
\newblock Journal of Fluid Mechanics \textbf{672}, 245--267 (2011)

\bibitem{duan2011direct}
Duan, L., Martin, M.: {Direct numerical simulation of hypersonic turbulent
  boundary layers. Part 4. Effect of high enthalpy}.
\newblock Journal of Fluid Mechanics \textbf{684}, 25 (2011)

\bibitem{Fu2020Direct}
Fu, L., Karp, M., Bose, S.T., Moin, P., Urzay, J.: {Shock-induced heating and
  transition to turbulence in a hypersonic boundary layer}.
\newblock Journal of Fluid Mechanics \textbf{909}, A8 (2021)

\bibitem{fu2013rans}
Fu, S., Wang, L.: {RANS modeling of high-speed aerodynamic flow transition with
  consideration of stability theory}.
\newblock Progress in Aerospace Sciences \textbf{58}, 36--59 (2013)

\bibitem{gaitonde1993calculations}
Gaitonde, D., Shang, J.: {Calculations on a double-fin turbulent interaction at
  high speed}.
\newblock In: 11th Applied Aerodynamics Conference, pp. AIAA--93--3432--CP
  (1993)

\bibitem{gaitonde1995structure}
Gaitonde, D., Shang, J., Visbal, M.: {Structure of a double-fin turbulent
  interaction at high speed}.
\newblock AIAA journal \textbf{33}(2), 193--200 (1995)

\bibitem{georgiadis2014status}
Georgiadis, N.J., Yoder, D.A., Vyas, M.A., Engblom, W.A.: {Status of turbulence
  modeling for hypersonic propulsion flowpaths}.
\newblock Theoretical and Computational Fluid Dynamics \textbf{28}(3), 295--318
  (2014)

\bibitem{gottlieb2001}
Gottlieb, S., Shu, C.W., Tadmor, E.: {Strong stability-preserving high-order
  time discretization methods}.
\newblock SIAM review \textbf{43}(1), 89--112 (2001)

\bibitem{hader2019direct}
Hader, C., Fasel, H.F.: {Direct numerical simulations of hypersonic
  boundary-layer transition for a flared cone: fundamental breakdown}.
\newblock Journal of Fluid Mechanics \textbf{869}, 341--384 (2019)

\bibitem{helmer2012three}
Helmer, D., Campo, L., Eaton, J.: {Three-dimensional features of a Mach 2.1
  shock/boundary layer interaction}.
\newblock Experiments in fluids \textbf{53}(5), 1347--1368 (2012)

\bibitem{huang2019assessment}
Huang, J., Bretzke, J.V., Duan, L.: {Assessment of Turbulence Models in a
  Hypersonic Cold-Wall Turbulent Boundary Layer}.
\newblock Fluids \textbf{4}(1), 37 (2019)

\bibitem{huang1995compressible}
Huang, P., Coleman, G., Bradshaw, P.: {Compressible turbulent channel flows:
  DNS results and modelling}.
\newblock Journal of Fluid Mechanics \textbf{305}, 185--218 (1995)

\bibitem{iyer2019analysis}
Iyer, P.S., Malik, M.R.: Analysis of the equilibrium wall model for high-speed
  turbulent flows.
\newblock Physical Review Fluids \textbf{4}(7), 074604 (2019)

\bibitem{kawai2012wall}
Kawai, S., Larsson, J.: {Wall-modeling in large eddy simulation: Length scales,
  grid resolution, and accuracy}.
\newblock Physics of Fluids \textbf{24}(1), 015105 (2012)

\bibitem{kawai2013dynamic}
Kawai, S., Larsson, J.: Dynamic non-equilibrium wall-modeling for large eddy
  simulation at high reynolds numbers.
\newblock Physics of Fluids \textbf{25}(1), 015105 (2013)

\bibitem{kussoy1992intersecting}
Kussoy, M., Horstman, K.: {Intersecting shock-wave/turbulent boundary-layer
  interactions at Mach 8.3}.
\newblock NASA Ames Research Center Technical Report, NASA-TM-103909  (1992)

\bibitem{lakebrink2019}
Lakebrink, M.T., Mani, M., Rolfe, E.N., Spyropoulos, J.T., Philips, D.A., Bose,
  S.T., Mace, J.L.: Toward improved turbulence-modeling techniques for
  internal-flow applications.
\newblock AIAA Paper 2019-3703  (2019)

\bibitem{larsson2016large}
Larsson, J., Kawai, S., Bodart, J., Bermejo-Moreno, I.: {Large eddy simulation
  with modeled wall-stress: recent progress and future directions}.
\newblock Mechanical Engineering Reviews \textbf{3}(1), 15--00418 (2016)

\bibitem{lehmkuhl2018}
Lehmkuhl, O., Park, G.I., Bose, S.T., Moin, P.: Large-eddy simulation of
  practical aeronautical flows at stall conditions.
\newblock Proceedings of the 2018 Summer Program, Center for Turbulence
  Research, Stanford University pp. 87--96 (2018)

\bibitem{lozano2020}
Lozano-Dur\'an, A., Bose, S.T., Moin, P.: Prediction of trailing edge
  separation on the {NASA} {J}uncture {F}low using wall-modeled {LES}.
\newblock AIAA Paper 2020-1776  (2020)

\bibitem{mani2009suitability}
Mani, A., Larsson, J., Moin, P.: {Suitability of artificial bulk viscosity for
  large-eddy simulation of turbulent flows with shocks}.
\newblock Journal of Computational Physics \textbf{228}(19), 7368--7374 (2009)

\bibitem{mettu2018wall}
Mettu, B.R., Subbareddy, P.K.: {Wall modeled LES of compressible flows at
  non-equilibrium conditions}.
\newblock AIAA Paper 2018-3405  (2018)

\bibitem{muto2019equilibrium}
Muto, D., Daimon, Y., Shimizu, T., Negishi, H.: {An equilibrium wall model for
  reacting turbulent flows with heat transfer}.
\newblock International Journal of Heat and Mass Transfer \textbf{141},
  1187--1195 (2019)

\bibitem{narayanswami1993numerical}
Narayanswami, N., Horstman, C., Knight, D.: {Numerical Simulation of Crossing
  Shock/turbulent Boundary Layer Interaction at Mach 8.3: Comparison of Zero
  and Two-equation Turbulence Models}.
\newblock AIAA paper 93-0779 (1993)

\bibitem{narayanswami1993investigation}
Narayanswami, N., Knight, D., Horstman, C.: {Investigation of a hypersonic
  crossing shock wave/turbulent boundary layer interaction}.
\newblock Shock Waves \textbf{3}(1), 35--48 (1993)

\bibitem{patel2015semi}
Patel, A., Peeters, J.W., Boersma, B.J., Pecnik, R.: {Semi-local scaling and
  turbulence modulation in variable property turbulent channel flows}.
\newblock Physics of Fluids \textbf{27}(9), 095101 (2015)

\bibitem{rumsey2010compressibility}
Rumsey, C.L.: {Compressibility considerations for kw turbulence models in
  hypersonic boundary-layer applications}.
\newblock Journal of Spacecraft and Rockets \textbf{47}(1), 11--20 (2010)

\bibitem{sandham2014transitional}
Sandham, N., Sch{\"u}lein, E., Wagner, A., Willems, S., Steelant, J.:
  {Transitional shock-wave/boundary-layer interactions in hypersonic flow}.
\newblock Journal of Fluid Mechanics \textbf{752}, 349--382 (2014)

\bibitem{souverein2010effect}
Souverein, L.J., Dupont, P., Debieve, J.F., Dussauge, J.P., Van~Oudheusden,
  B.W., Scarano, F.: {Effect of interaction strength on unsteadiness in
  shock-wave-induced separations}.
\newblock AIAA journal \textbf{48}(7), 1480--1493 (2010)

\bibitem{vreman2004eddy}
Vreman, A.: {An eddy-viscosity subgrid-scale model for turbulent shear flow:
  Algebraic theory and applications}.
\newblock Physics of Fluids \textbf{16}(10), 3670--3681 (2004)

\bibitem{wangmoin2002}
Wang, M., Moin, P.: {Dynamic wall modeling for large-eddy simulation of complex
  turbulent flows}.
\newblock Physics of Fluids \textbf{14}(7), 2043--2051 (2002)

\bibitem{wu2017inflow}
Wu, X.: Inflow turbulence generation methods.
\newblock Annual Review of Fluid Mechanics \textbf{49}, 23--49 (2017)

\bibitem{yang2017aerodynamic}
Yang, X., Urzay, J., Bose, S., Moin, P.: {Aerodynamic heating in wall-modeled
  large-eddy simulation of high-speed flows}.
\newblock AIAA Journal pp. 731--742 (2017)

\bibitem{yang2018semi}
Yang, X.I., Lv, Y.: {A semi-locally scaled eddy viscosity formulation for LES
  wall models and flows at high speeds}.
\newblock Theoretical and Computational Fluid Dynamics \textbf{32}(5), 617--627
  (2018)

\end{thebibliography}

\end{document}